\documentclass[aps,prd,nofootinbib,notitlepage,superscriptaddress,eprint]{revtex4-1}
\usepackage{graphicx}
\usepackage{amsmath,amssymb,amsfonts}
\usepackage{braket}
\usepackage{float}
\usepackage{xcolor}
\usepackage{soul}
\usepackage{rotating}
\usepackage{multirow}
\usepackage{mathtools}
\usepackage[normalem]{ulem}
\usepackage{comment}
\usepackage{cprotect}
\usepackage{xspace}

\usepackage{makecell}

\usepackage{hyperref}
\hypersetup{
	colorlinks=true,
	linkcolor=blue,
	citecolor=blue,
	urlcolor = blue,
}

\def\nn{\nonumber}
\def\pa{{\partial}}

\def\l{\left}
\def\r{\right}
\def\d{{\rm d}}
\def\th{\tilde{H}}
\def\vx{{\bf x}}
\def\vk{{\bf k}}

\def\th{\tilde{H}}

\def\Mp{M_{\scriptscriptstyle{\rm Pl}}}
\newcommand{\sR}[1][3]{{}^{(#1)}\!R}

\begin{document}

\title{Alternative approach to the Starobinsky model for inflation scenarios}

\author{Masud Chaichian}
\email{masud.chaichian@helsinki.fi}
\affiliation{Department of Physics, University of Helsinki, P.O. Box 64, FI-00014 Helsinki, Finland}
\affiliation{Helsinki Institute of Physics, P.O. Box 64, FI-00014 University of Helsinki, Finland}
\author{Amir Ghal'e}
\email{ghalee@tafreshu.ac.ir}
\affiliation{Department of Physics, Tafresh University,
P. O. Box 39518-79611, Tafresh, Iran}
\author{Markku Oksanen}
\email{markku.oksanen@helsinki.fi}
\affiliation{Department of Physics, University of Helsinki, P.O. Box 64, FI-00014 Helsinki, Finland}
\affiliation{Helsinki Institute of Physics, P.O. Box 64, FI-00014 University of Helsinki, Finland}

\begin{abstract}
The $R+R^2$ model of gravity with the corresponding shallow potential in the Einstein frame is consistent with the observations. Recently, many efforts have been made to generalize the $R+R^2$ (Starobinsky) model of inflation or use other shallow potentials to construct a model for the early Universe. We revise the question about the shallow potential. We propose a model in which the Starobinsky model can emerge through a dynamical mechanism. We show that the absence of ghost modes results to constraints on the parameters of the Starobinsky model.
We obtain the scalar spectral index and the tensor-to-scalar ratio of the extended model and study the three-point correlation function of the curvature perturbation to estimate the primordial non-Gaussianities of the proposed model.
\end{abstract}
\maketitle

\section{Introduction}
The recent measurements of the anisotropies of the cosmic microwave background (CMB) by the \textit{Planck} mission \cite{{Planck1},{Planck2},{Planck3}} provide more accurate values for cosmological parameters.
The data support the cosmic inflation paradigm, which involves a period of accelerated expansion in the very early Universe that was originally proposed in \cite{{stra},{kaz},{chib},{gu},{lin1},{st}}.
Confirmation of the inflation scenario for the early Universe could come from the detection of the gravitational waves as $B$ modes in CMB. The expectation for such gravitational waves and the data have provided significant constraints on the models of inflation \cite{Planck3}.
For example, the data are not consistent with some popular models such as quadratic chaotic inflation \cite{Planck3}.
On the other hand, the observations confirm the existence of the dark sectors in the Universe \cite{Planck3}. However, the nature of the dark sectors remains as an open question in cosmology and particle physics \cite{apr}.

\begin{figure}[ht]
\includegraphics[width=0.85\textwidth]{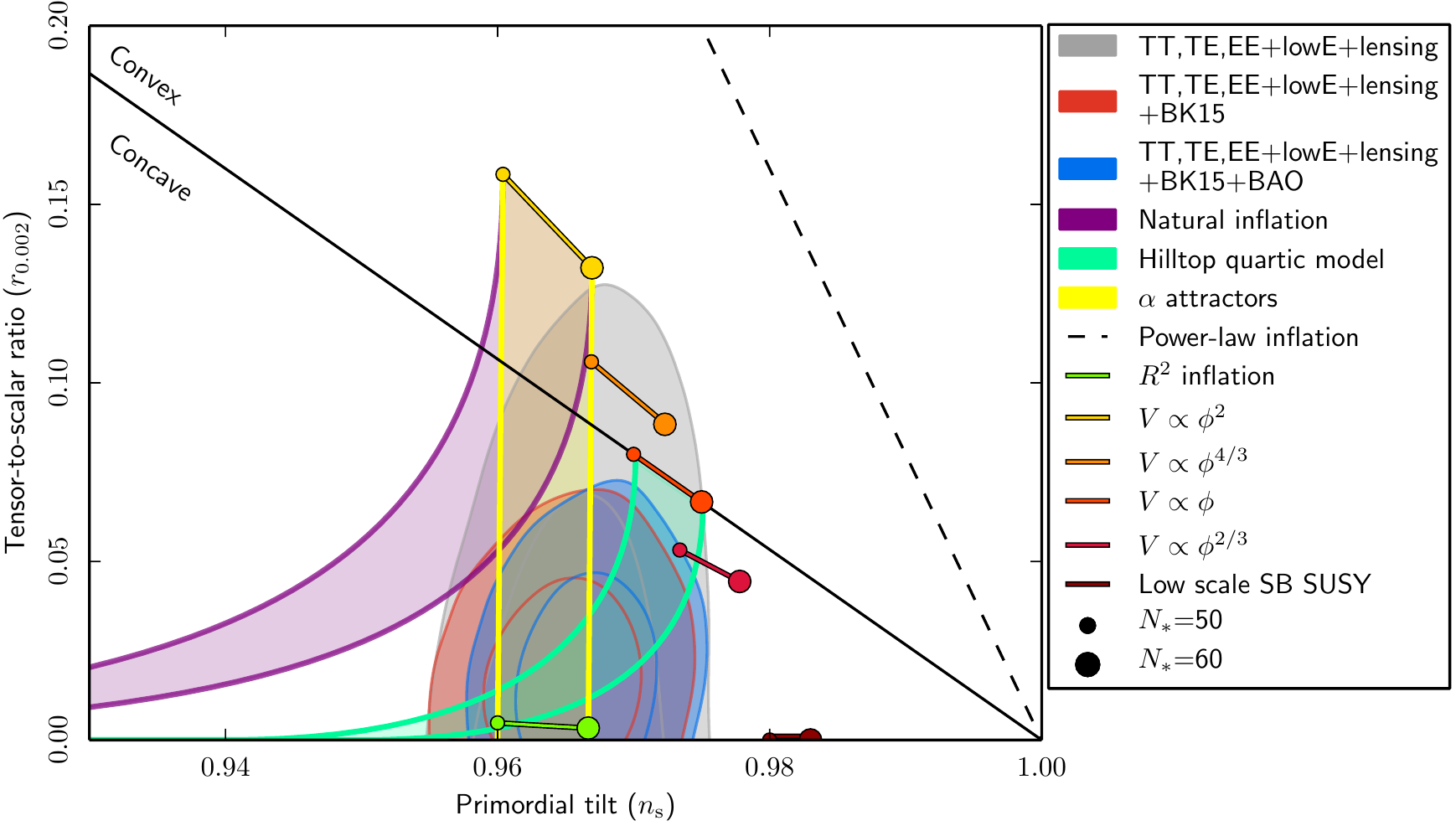}
\caption{Marginalized joint 68\,\% and 95\,\% confidence level regions for the scalar spectral index $n_s$ and the tensor-to-scalar ratio $r$ at the wavenumber $k=0.002\,\mathrm{Mpc}^{-1}$ of the perturbations.
Grey regions are from the Planck data alone, red regions are in combination
with the measurement of the CMB $B$-mode polarization angular power spectrum by the BICEP2/Keck Array collaboration (BK15), and
blue regions further include measurements of the baryon acoustic oscillation (BAO) scale.
The theoretical predictions of selected models are plotted for the number of \textit{e}-folds of the inflationary expansion in the range $50\le N_{*}\le60$.
The Starobinsky model is included as ``$R^2$~inflation''.
Figure courtesy of Planck Collaboration~\cite{Planck3}, licensed under \href{https://creativecommons.org/licenses/by/4.0}{CC BY 4.0}.}
\label{fig0}
\end{figure}

The Starobinsky model (or the $R+R^2$ model) is described by a Lagrangian \cite{Sakharov}
\begin{equation}\label{i-0}
\mathcal{L}_{s}=\sqrt{-g}\left[\frac{\Mp^2 R}{2}+ f_{s}R^2\right],
\end{equation}
where $\Mp$ is the reduced Planck mass and $f_s$ is a dimensionless constant. This model is in good agreement with the observations~\cite{Planck3}. Fig.~\ref{fig0} shows the agreement (and disagreement) of a selection of inflationary models with the observations.
It is worthwhile to mention that by introducing a mass scale, $M$, one can parametrize $f_s$  as $f_s=\frac{\Mp^2}{12M^2}$ \cite{Planck3}. Comparing the Einstein-Hilbert Lagrangian with the Starobinsky model, the latter model has an additional scalar degree of freedom which produces the inflationary era.\footnote{As a side remark it seems that in the eternal inflation theorem \cite{Borde:2001nh} (see also \cite{Linde:2007fr})  together with an isotropic expansion, there is an implicit assumption on the  validity of the Strong Equivalence Principle (SEP) of the acting gravitational theory therein. In that case, the Starobinsky model for inflation, being a modified gravity theory does not have the SEP and thus does not satisfy the requirements of the eternal inflation.}

It is worth noting that there are no ghost modes in the Starobinsky model \cite{stelle}. For this reason we do not include the other quadratic curvature terms, i.e. the square of the Ricci tensor $R_{\mu\nu}R^{\mu\nu}$ and the square of the Weyl tensor $C_{\mu\nu\alpha\beta}C^{\mu\nu\alpha\beta}$, which would involve the problem with ghosts \cite{KOT}.

The Lagrangian \eqref{i-0} represents the Starobinsky model in the Jordan frame \cite{jordan}, for which the cosmological perturbations were obtained in Ref. \cite{mukhanov1}.
It is convenient to use a conformal transformation to represent the Starobinsky model in the so-called the Einstein frame \cite{mukhanov1}. For this purpose we may perform a conformal transformation as $\tilde{g}_{\mu\nu}=(1+\frac{4f_s}{\Mp^2}R)g_{\mu\nu}$ and define a scalar field $\phi$ as
\begin{equation}
R=\frac{\Mp^2}{2}\left(e^{\frac{2\phi}{\sqrt{6}\Mp^2}}-1\right).
\end{equation}
Then, the Starobinsky model takes the following form
\begin{equation}\label{ei0}
\mathcal{L}_{E}=\sqrt{-\tilde{g}}\left[\frac{\Mp^2}{2}\tilde{R}-\frac{1}{2}\tilde{g}^{\mu\nu}\partial_\mu\phi\partial_\nu\phi-V(\phi)\right],
\end{equation}
where $\tilde{R}$ is the Ricci scalar that is constructed from the metric $\tilde{g}_{\mu\nu}$ and the potential is defined as
\begin{equation}
V(\phi)=\frac{\Mp^4}{16f_s}\left[1-\exp\left(\frac{-2\phi}{\sqrt{6}\Mp}\right)\right]^2.
\end{equation}
Let us clarify some reasons why one usually prefers to use the Einstein frame and criticize some of them. The first reason is that the dynamics of scalar fields are well known in particle physics. So, there exist many inflationary theories which draw motivations from particle physics. However, the shape of $V(\phi)$ shows that it is a highly nontrivial task to obtain such a potential from particle physics. The second reason is that the study of the reheating era can be done with the standard tools of particle physics. However, from the observations, the identity and interactions of the dark sector with the Standard Model are unknown. The Einstein frame is very useful if we neglect other (matter) fields in the Universe. Otherwise, the conformal transformation induces non-minimal interaction with the matter fields.
The last but not least reason is that Ref. \cite{Planck3} gives the constraints on inflationary cosmological models in the Einstein frame. For example, Ref. \cite{Planck3} gives a lower bound on $f_s$ as $f_s\geq 3\times 10^{9}$ \cite{{gr2n},{gr3}}.\footnote{It is instructive to define $\theta=\frac{1}{f_s}$ and $V(\phi)=\frac{\theta\Mp^4}{16}\left[1-\exp\left(\frac{-2\phi}{\sqrt{6}\Mp}\right)\right]^2$. Now we can ask why $\theta$ has such a small value and quest a dynamical mechanism to explain it. This question is similar to the question on the strong CP-violation term in QCD for which the Peccei-Quinn mechanism was proposed \cite{peccie}.
Analogously to the strong CP problem in QCD, the present attempt aims to propose a mechanism which would make $\theta=\frac{1}{f_s}$ sufficiently small. Note that a possible solution for $\theta$-term is the Peccei-Quinn mechanism in which a scalar field is introduced to answer the question.}

The shape of $V(\phi)$ with the lower bound on $f_s$ raises some questions about the Starobinsky model. When $f_s\gg 1$, $V(\phi)$ is a  shallow potential. Since this potential is consistent with the observations, many authors have considered the question ``\textit{why such a shallow potential?}'' and proposed modified versions of $V(\phi)$ \cite{{fr},{gr2},{re3},{re},{od}}.

In this paper we propose an extension of the Starobinsky model in the Jordan frame. The first reason for using the Jordan frame is that we wish to turn the question ``\textit{why the shallow potential?}'' to ``\textit{why $f_s$ has this lower bound?}''. We will show that it is possible to construct a model from where the Starobinsky model can emerge. We will show that the lower bound on $f_s$ results to the absence of the ghost modes and the tachyonic instability in the proposed model. As we will see, our results provide an alternative interpretation of the Starobinsky model.

The organization of this paper is as follows. In Sec.~\ref{sec2} we study the background cosmology of the model. Sec.~\ref{sec3} is devoted to the study of the linear perturbation equations, and we derive the scalar spectral index and the tensor-to-scalar ratio of the model.  Furthermore, we study ghost modes and tachyonic instability.
In Sec.~\ref{sec4} we derive the corresponding cubic action for scalar metric perturbations in order to study the primordial non-Gaussianities of the model. Sec.~\ref{sec5} contains conclusions.

\section{The model}\label{sec2}
In this section we propose the model and then discuss the background cosmology of our model.
We introduce a scalar field $\varphi$ and consider the following gravitational action
\begin{equation}\label{i-1}
S=\int d^{4}x\sqrt{-g}\left[\frac{\Mp R}{2}+ f(X) R^2\right],\quad X\equiv\frac{-\partial_\mu \varphi\partial^\mu\varphi}{2M^4},
\end{equation}
where $f$ is a function and $M$ is a constant of dimension mass.
Varying the action (\ref{i-1}) with respect to the metric, gives
\begin{equation}\label{s1_2}
\begin{split}
R_{\mu\nu}-\frac{1}{2}Rg_{\mu\nu}=&\ 2\Mp^{-2}\bigg(\frac{\pa_\mu\varphi\pa_{\nu}\varphi}{2M^4}f_X R^2
-2RfR_{\mu\nu}\\
&+2\pa_{\mu}R\pa_{\nu}f+2\pa_\nu R\pa_\mu f
+2R\nabla_{\mu}\nabla_{\nu} f\\
&+2f\nabla_\mu\nabla_\nu R
-2f\nabla^2Rg_{\mu\nu}-2R\nabla^2fg_{\mu\nu}\\
&-4\pa^\alpha R\pa_\alpha f g_{\mu\nu}+\frac{fR^2}{2}g_{\mu\nu}\bigg),
\end{split}
\end{equation}
where $f=f(X)$ and $f_X=\frac{df}{dX}$.
Varying the action (\ref{i-1}) with respect to $\varphi$, gives
\begin{equation}\label{s1_8}
\nabla_\mu j^{\mu}=\partial_\mu j^\mu=0,\quad
j^{\mu}=\sqrt{-g}R^2f_X\partial^{\mu}\varphi,
\end{equation}
which reflects the fact that the action \eqref{s1_2} depends on $\varphi$ only through its derivatives $\partial_\mu\varphi$ and is thus invariant under constant shifts of the scalar field, $\varphi(x)\rightarrow\varphi(x)+\alpha$, with $\alpha=\text{const}$.
The corresponding conserved quantity is $Q=\int_{\Sigma_t}d^3x\sqrt{h}NR^2f_X\partial^{0}\varphi$, where $h$ is the determinant of the induced metric on the spatial hypersurface $\Sigma_t$ at time $t$ and $N$ is the lapse function (see Eq.~\eqref{s2_16}).
The boundary conditions (or asymptotic conditions) are assumed to be such that the following integral over the boundary of $\Sigma_t$ vanishes, $\int_{\partial\Sigma_t}d^2x\sqrt{\gamma}NR^2f_Xr_i\partial^i\varphi$, where $\gamma$ is the determinant of the induced metric on the boundary and $r_i$ is the unit normal to the boundary.

In this paper our aim has been to provide a new perspective for the Starobinsky model as an emergent model, while being led by the results of the Planck observations \cite{{Planck1},{Planck2},{Planck3}}
to impose the shift symmetry for the new  degree of freedom in the model.
After those observations,  it turns out that only a field  with a potential of plateau type, such as $V=V_0-e^{-\alpha\phi}$, is consistent with the data.
Also, according to the data, it is necessary that such potentials approach a constant value at the inflationary phase, which means that such a potential has to have an approximate shift symmetry at the inflationary phase (see the Fig.~\ref{fig0} for the constraints on various models.)
At this point, it would be a motivation to choose the shift symmetry as a guideline for construction of a model from the very beginning.
The shift symmetry has previously been used in cosmology. Indeed, using the shift symmetry in the cosmological models where a scalar field is minimally coupled to gravity, results in a reduced speed of sound and consequently a significant non-Gaussianity. Such facts  lead us to consider a nonminimally coupled scalar field with the shift symmetry as a plausible model.

We have used the aforementioned shift symmetry as a guideline in obtaining the action {\eqref{i-1}}, which reproduces the behavior of the Starobinsky model during inflation and provides motivation for the large value of the coupling $f_s$. Thus, it is essential that the action does not involve a potential term for the field $\varphi$ or interaction terms between $\varphi$ and curvature. Among the crucial criteria for the model are to ensure that the tensor-to-scalar ratio of the dimensionless power spectra of the perturbations is sufficiently small, the speed of sound of the perturbations is not changed significantly, and that the primordial non-Gaussianity remains negligible.

It should be mentioned that, at first glance, the action \eqref{i-1} is similar to the action considered in Refs. \cite{{noh1},{noh2}}. However, inspection of Refs. \cite{{noh1},{noh2}} shows that although the authors began with a general action, they did not consider our model.\footnote{In Ref. \cite{noh1}, the authors begin with a gravitational Lagrangian that depends generally on $R$, $\varphi$ and $X$. However, before Eq.~(10) of Ref.~\cite{noh1} they clarify that the authors focused on certain cases, where the Lagrangian density $\mathcal{L}$ is such that $\frac{\partial(\mathcal{L}/\sqrt{-g})}{\partial R}=F(\varphi)$ with $F$ some function of $\varphi$. Those cases do not include our model, where $F$ depends now on $R$ and $X$. Similarly, in Ref. \cite{noh2} a general action is introduced as Eq.~(74), but just after Eq. (83) in Ref. \cite{noh2}, they focused on specific models which are not the same as our model.}

Theories that couple the derivatives of a scalar field $\Phi$ to the first power of curvature, e.g. as the terms $G^{\mu\nu}\partial_\mu\Phi\partial_\nu\Phi$ and $R\partial_\mu\Phi\partial^\mu\Phi$, where $G_{\mu\nu}$ is the Einstein tensor, have also been proposed and studied previously, particularly regarding their implications on cosmology. In particular, it has been proposed that the scalar field in such a theory is the Higgs field of the standard model of particle physics {\cite{Germani:2010gm}}. Indeed, this has been an appealing way to produce inflation with a known scalar field, providing an interesting new connection between particle physics and cosmology. Also the cosmological perturbations in this model were studied in {\cite{Germani:2010ux}}.
This new version of Higgs inflation is an alternative to the widely studied Higgs inflation, where the scalar field is coupled to curvature as $|\Phi|^{2}R$, see {\cite{Cervantes-Cota:1995ehs}}, {\cite{Bezrukov:2007ep}} and for a review {\cite{Rubio:2018ogq}}. A much studied issue with Higgs inflation has been the potential violation of unitarity during inflation; for recent studies see {\cite{Antoniadis:2021axu}} and {\cite{Ito:2021ssc}}. The Higgs inflation model together with the Starobinsky term $R^2$ has also been studied, e.g. in {\cite{He:2018gyf}} and {\cite{Antoniadis:2018ywb}}, and it has been argued that the large value of the constant $f_s$ of the term $R^2$ would be induced by the quantum effects due to the coupling $|\Phi|^{2}R$ {\cite{Calmet:2016fsr}}.

We also remark that an extension of the Starobinsky model \eqref{i-0}, where the scalar curvature terms are coupled to functions of a scalar field, has been explored in \cite{Kaneda:2015jma}. Such models, and more generally models with arbitrary couplings of a scalar field and the scalar curvature, can extend or combine the properties of Starobinsky or $f(R)$ gravity and scalar field theories.

In Appendix~\ref{appendix1}, we show that the action \eqref{i-1} is conformally equivalent to the following action
\begin{equation}\label{reei0}
S_{E}=\int d^4x\sqrt{-\tilde{g}}\left[\frac{\Mp^2}{2}\tilde{R}-\frac{1}{2}\tilde{g}^{\mu\nu}\partial_\mu\phi\partial_\nu\phi-V(\phi,\varphi)\right],
\end{equation}
where $\tilde{g}_{\mu\nu}=(1+\frac{4f(X)}{\Mp^2}R)g_{\mu\nu}$ and
\begin{equation}\label{revisedei1}
V(\phi,\varphi)=\frac{\Mp^4}{16f\left(-\frac{1}{2M^4}e^{(\frac{2\phi}{\sqrt{6}Mp})}\tilde{g}^{\mu\nu}\partial_{\mu}\varphi\partial_\nu\varphi\right)}\left[1-\exp\left(\frac{-2\phi}{\sqrt{6}\Mp}\right)\right]^2.
\end{equation}

The main result in this section is that the scalar field in \eqref{i-1} moves towards a minimum of $f(X)$ and is condensed during inflation. The condensed phase for the scalar field shows that we have a preferred observer, which is not a new phenomena in the cosmological context \cite{{armendariz},{arka}}.

As we pointed out, we would like to provide some reasons for the lower bound on $f_s$. It is clear that when we consider a constant value for $f(X)$ as $f(X)=f_s$, our results must reduce to the corresponding results for the Starobinsky model. So, for a cross-check of our equations, one can use \cite{re} with a constant function $f(X)$.
Also, $\varphi$ could be considered as a scalar field for the dark sectors.

\subsection{Background cosmology}
We use the flat Friedmann-Robertson-Walker (FRW) metric as the cosmological background metric
\begin{equation}\label{s1_1}
ds^2=-dt^2+a^2\delta_{ij}dx^idx^j,
\end{equation}
where $a=a(t)$ is the scale factor from which the Hubble parameter is defined as
$H=\frac{\dot{a}}{a}$, where the dot denotes the time derivative.
Using the FRW metric, the time-time  and space-space components of Eq. (\ref{s1_2}) can be obtained as
\begin{equation}\label{s1_3}
3H^2=6H\tilde{H}-\frac{2\Mp^{-2}\l[Xf_X R^2+\frac{fR^2}{2}\r]}{(1+4\Mp^{-2}Rf)}
\end{equation}
and
\begin{align}\label{s1_4}
-(3H^2+2\dot{H})=&4(\tilde{H}-H)H+4(\tilde{H}-H)^2\\\nn
&+2(\dot{\tilde{H}}-\dot{H})-\frac{\Mp^{-2}fR^2}{(1+4\Mp^{-2}Rf)},
\end{align}
respectively, where
\begin{equation}\label{s1_5}
\tilde{H}=H+\frac{2\Mp^{-2}}{(1+4\Mp^{-2}Rf)}\frac{d}{dt}(Rf).
\end{equation}
From Eqs. (\ref{s1_3}) and (\ref{s1_4}), it follows that
\begin{equation}\label{s1_6}
\dot{\tilde{H}}=H(\tilde{H}-H)-2(\tilde{H}-H)^2-\frac{\Mp^{-2}Xf_X R^2}{(1+4\Mp^{-2}Rf)}.
\end{equation}
Hence, from Eq. (\ref{s1_5}) it follows that as far as $1+4\Mp^{-2}Rf>0 $, one can define the following variable
\begin{equation}\label{s1_7}
\tilde{a}=a(1+4\Mp^{-2}Rf)^{\frac{1}{2}}
\end{equation}
and that
\begin{equation}
\tilde{H}=\frac{\dot{\tilde{a}}}{\tilde{a}}.
\end{equation}
The scalar field Eq. \eqref{s1_8} with $\varphi=\varphi(t)$ and the FRW metric is obtained as
\begin{equation}
\frac{d}{dt}\left(a^3R^2f_X\dot\varphi\right)=0,
\end{equation}
which leads us to the following equation
\begin{equation}\label{s1_9}
\frac{\dot{X}}{HX}=-6\l(1+\frac{2}{3}\frac{\dot{R}}{HR}\r)\frac{f_X}{f_X+2Xf_{XX}},
\end{equation}
where $f_{XX}=\frac{d^2f}{dX^2}$.

\subsection{Inflationary era}
Before we study the inflation scenario for the proposed model, let us bear in mind that the de Sitter solution does not exist in the Starobinsky model. The de Sitter solution describes a universe that inflates eternally, which is not suitable for the early Universe. For the early Universe, we just need a solution
that remains close to the de Sitter solution, i.e. we need the quasi-de Sitter solution. It is easy to show that for the proposed model, the de Sitter solution does not exist. To clarify this issue, note that the de Sitter solution is characterized by a constant value of the Hubble parameter as $H_{ds}$, which determines the Ricci scalar as $R_{ds}=12H^2_{ds}$. Inserting these into Eqs. \eqref{s1_3}--\eqref{s1_6} results in $H_{ds}=\tilde{H}_{ds}=0$, which shows that, for the proposed model, the de Sitter solution does not exist. As we pointed out, we just need a quasi-de Sitter solution for the early Universe.
The slow-roll conditions are characterized by
the Hubble flow functions (HFFs), $\epsilon_i$, as
\begin{equation}\label{cond}
 \epsilon_1=-\frac{\dot{H}}{H^2}\ll1,\quad \epsilon_{n+1}=\frac{\dot{\epsilon}_n}{H\epsilon_n}\ll1\quad (n\geq 1).
\end{equation}
According to the \textit{Planck} results \cite{Planck3}, we have
\begin{align}\label{pl-cond}
&\epsilon_1<0.0063 \qquad\quad (0.0039) \qquad\qquad (95\%~\text{CL}),   \nn\\
&\epsilon_2=0.030_{-0.005}^{+0.007} \quad (0.031 \pm 0.005) \quad (68\% ~\text{CL}),
\end{align}
where the values in parentheses additionally take into account the data from the BICEP2/Keck Array, and CL is the confidence level.
Furthermore, for any time-dependent quantity such as $X(t)$ we have to impose $\frac{\dot{X}}{HX}\ll1$.

It should be mentioned that to report the above data, the Einstein frame is used in \cite{Planck3}. In this work, we will use the Jordan frame \cite{re}. Although we just need to know that during inflation we have $\epsilon_i\ll1$, in a few places we feel that the numerical values for $\epsilon_i$ help us to clarify some issues. In such cases, we will take $\epsilon_1=4\times10^{-3}$. In other words, we make the strong assumption that during inflation the order of magnitude of $\epsilon_i$ are the same in the both frames.\footnote{The Einstein and Jordan frame are related by a conformal transformation. There is no physical principle that shows invariance of physical quantities under the conformal transformation. For an example, see Eq. (\ref{report_te}).}

Although the slow-roll conditions can be regarded as initial conditions, they must remain valid during inflation. Hence the consistency of the conditions in Eq. (\ref{cond}) with the dynamics of the model must be investigated.
Using $R=12H^2+6\dot{H}$, Eq. (\ref{s1_9}) can be rewritten as
\begin{align}\label{s1_11}
\frac{dX}{d\cal{N}}=\frac{-6Xf_X\l[1-\frac{4}{3}\epsilon_1\l(1+\frac{\epsilon_2}{2(2-\epsilon_1)}\r)\r]}{f_X+2Xf_{XX}},
\end{align}
where ${d\cal{N}} = Hdt=d\ln a$, which measures the number of \textit{e}-folds ${\cal{N}}$ of inflationary expansion.
Note that the above equation is exact, and in the above equation and all differential equations for $\epsilon_i$ we have $\epsilon_i=\epsilon_i(t)$.

In what follows, we make the ansatz that
\begin{equation}
\text{$X$ is moving to a minimum of $f(X)$.}
\end{equation}
It should be noted that it is not obvious that the ansatz is correct and leads to a consistent solution. So, we have to check the ansatz at the end of the calculations. We also have to check that if one consider $\epsilon_i\ll1$ as the initial value at the beginning of the era of inflation, then $\epsilon_i$ will remain small during inflation.

Now, we want to study the model close to the minimum of $f(X)$ which can be represented as $X=1$. Note that this expansion is a nontrivial step. We must show that this assumption leads us to a well-defined behaviour for the model. As we will show, our results support this assumption.
Although our discussions are general, in what follows one can consider $f(X)=f_s+\frac{{\cal{F}}}{2}(X-1)^2+\ldots$. As an example, if we take the ``Higgs-like'' shape as $f(X)=a_1-a_2 X+a_3X^2$ then $f(X)$ has a minimum at $X_*=\frac{a_2}{2a_3}$ . Then by using $f_s\equiv a_1-\frac{a_2^2}{4a_3}$ and ${\cal{F}}\equiv2a_3$, it turns out that $f(X)=f_s+\frac{{\cal{F}}}{2}(X-X_*)^2$. It is always possible to take $X_*=1$ at the background level.

Expanding Eq. (\ref{s1_11}) to first order in $(X-1)$ gives
\begin{equation}\label{s1_12}
\frac{dX}{d\cal{N}}=-3K(\epsilon_1,\epsilon_2)(X-1),
\end{equation}
where
\begin{equation}\label{s1_13}
K(\epsilon_1,\epsilon_2)\equiv1-\frac{4}{3}\epsilon_1\l(1+\frac{\epsilon_2}{2(2-\epsilon_1)}\r).
\end{equation}
To determine behaviour of $K(\epsilon_1,\epsilon_2)$, expanding Eq. (\ref{s1_3}) to first
order in $X-1$  results in
\begin{align}\label{s1_14}
\frac{1}{24f\Mp^{-2}\epsilon_1 H^2(2-\epsilon_1)}-\frac{3}{2}=&\frac{\dot{\epsilon}_1}{H\epsilon_1(2-\epsilon_1)}\\\nn
&+(X-1)\frac{f_{XX}(2-\epsilon_1)}{f\epsilon_1},
\end{align}
where we have used $R=6H^2(2-\epsilon_1)$ which results in $\dot{R}=-12H^3\epsilon_1(2-\epsilon_1)-6H^2\dot{\epsilon_1}$.
Note that, Eq. (\ref{s1_14}) is an exact equation around the minimum of $f(X)$ and we just write it in terms of $\epsilon_1$ .
Before investigate the inflationary era of the proposed model, let us clarify our motivations to study the model. For this goal, from Eq. (\ref{s1_14}) it turns out that for the Starobinsky model we have to take the following condition to have the slow-roll inflation
\begin{align}\label{s1_af1}
\frac{1}{24f_s\Mp^{-2}\epsilon_1 H^2(2-\epsilon_1)}-\frac{3}{2}=g_{s} \epsilon_1\ll1\\\nn
 (\text {for the Starobinsky model }),
\end{align}
where $g_s\equiv\frac{\epsilon_2}{\epsilon_1(2-\epsilon_1)}$.
The above equation can be rewritten as
\begin{equation}\label{s1_afff}
\frac{1}{4f_s\Mp^{-2}\epsilon_1 R}-\frac{3}{2}=g_s\epsilon_1\ll1  \hspace{.3cm}(\text {for the Starobinsky model }).
\end{equation}
Another parametrization of the above equation for the Starobinsky model is presented in Ref.~\cite{rep1}.
Since we have $\epsilon_{i}\ll1$ during inflation, Eq. (\ref{s1_afff}) gives $\Mp^{-2}Rf_s\gg1$ but the reverse is not true.
We also know from Ref. \cite{Planck3} that $\frac{H}{\Mp}<2.5\times10^{-5}$. Combining this with the data from \eqref{pl-cond} and Eq. \eqref{s1_af1} gives us a bound on $f_s$, which is shown in Table~\ref{table0}.
\begin{table}
\renewcommand{\arraystretch}{1.5}
    \caption{Bound on $f_s$ in the Starobinsky model. The bound can be obtained by using Eq. \eqref{s1_af1} and the Planck results that give $\frac{H}{\Mp}<2.5\times10^{-5}$ and \eqref{pl-cond}.}
    \begin{tabular}{ | c | c|}
    \hline
    $\epsilon_1$ & Bound on $f_s$  \\ \hline
    0.006 & $f_s\geq3.7\times10^9$  \\ \hline
    0.004 & $f_s\geq5.5\times10^9$  \\ \hline
    0.001 & $f_s\geq2.2\times10^{10}$  \\ \hline
    \end{tabular}\label{table0}
\end{table}

As in the Starobinsky model, we take the following condition as the necessary condition for inflation in our model
\begin{equation}\label{s1_af}
\frac{1}{24f\Mp^{-2}\epsilon_1 H^2(2-\epsilon_1)}-\frac{3}{2}\equiv g\epsilon_1\ll1,
\end{equation}
where $g\equiv\frac{\epsilon_2}{\epsilon_1(2-\epsilon_1)}+(X-1)\frac{f_{XX}(2-\epsilon_1)}{f\epsilon_1^2}$. As noted earlier, at the end of calculations, we have to check such assumptions.
From Eqs. (\ref{s1_14}) and (\ref{s1_af}) it follows that
\begin{equation}\label{1400}
  \frac{\dot{\epsilon}_1}{H\epsilon_1(2-\epsilon_1)}+(X-1)(2-\epsilon_1)\frac{f_{XX}}{f\epsilon_1}=g\epsilon_1.
  \end{equation}
Expanding Eq. (\ref{s1_6}) to first order in $(X-1)$  and writing the result in terms of $\epsilon_1$ gives
\begin{align}\label{s1_150000}
&\frac{\ddot{\epsilon}_1}{2H^2(2-\epsilon_1)}+\frac{\dot{\epsilon}_1}{2H(2-\epsilon_1)}(3-6\epsilon_1)\\\nn
&-\frac{3}{2}\frac{f_{XX}}{f}(X-1)(2-\epsilon_1)-3\epsilon_1^2\\\nn
&=-\frac{\epsilon_1}{24f\Mp^{-2}H^2(2-\epsilon_1)}.
\end{align}
Then inserting Eq. (\ref{s1_af}) into Eq. (\ref{s1_150000}) gives
\begin{align}\label{s1_15}
&\frac{\ddot{\epsilon}_1}{2H^2(2-\epsilon_1)}+\frac{\dot{\epsilon}_1}{2H(2-\epsilon_1)}(3-6\epsilon_1)\\\nn
&+(g-\frac{3}{2})\epsilon_1^2
-\frac{3}{2}\frac{f_{XX}}{f}(X-1)(2-\epsilon_1)=0.
\end{align}
From Eqs. (\ref{1400}) and (\ref{s1_15}), we can eliminate the terms which are proportional to $(X-1)$. This yields a differential equation for $\epsilon_1$ as
\begin{equation}\label{s1_16}
\frac{d^2\epsilon_1}{{d\cal{N}}^2}+(6-7\epsilon_1)\frac{d\epsilon_1}{d\cal{N}}-(3+g)\epsilon_1^2(2-\epsilon_1)=0,
\end{equation}
where $d{\cal{N}}=Hdt$ is used.

The above equation is an example of the Lienard's equation \cite{perko}. There exists a systematic procedure to study the Lienard's equation. By introducing an auxiliary variable $y$, it is easy to find the following representation for Eq. (\ref{s1_16})
\begin{equation}\label{s1_17}
 \begin{cases}
   \frac{d\epsilon_1}{d\cal{N}}=-6\epsilon_1+\frac{7}{2}\epsilon_1^2 +y,&\\
   \frac{dy}{d\cal{N}}=(3+g)\epsilon_1^2(2-\epsilon_1).&
  \end{cases}
\end{equation}
Eq. (\ref{s1_17}) is an autonomous system and has the following fixed points in the phase space of $(\epsilon_1,y)$,
\begin{equation}\label{s1_18}
(0,0),\quad (2,-2).
\end{equation}
During inflation, $\epsilon_1\ll 1$, we have to take the initial values around $(0,0)$.
To determine the behaviour of this fixed point, we linearize the system (\ref{s1_17}) around $(0,0)$ as
\begin{equation}\label{s1_19}
\begin{pmatrix}
\frac{d\epsilon_1}{d\cal{N}}&   \\
\frac{dy}{d\cal{N}} &
\end{pmatrix}=A_{*}\begin{pmatrix}
\epsilon_1 &   \\
y &
\end{pmatrix},
\end{equation}
where $A_{*}$ is the stability matrix around $(0,0)$, which has the following form
\begin{equation}\label{s1_20}
A_{*}=\begin{pmatrix}
-6 & 1  \\
0 & 0
\end{pmatrix}.
\end{equation}
The above matrix has two eigenvalues as $\lambda_1=-6$ and $\lambda_2=0$. Thus, $(0,0)$ is the stable fixed point.
Therefore, if we take $\epsilon_1\ll1$ as the initial condition, this result shows that $\epsilon_1$ remains small during inflation. Note that the stability of $(0,0)$ does not depend on $g$.
Eq. (\ref{s1_19}) can be solved as
\begin{equation}\label{s1_21}
\epsilon_1=C_1e^{-6\cal{N}}+C_2\quad\text{(close to $(0,0)$)}.
\end{equation}
where $C_1$ and $C_2$ are constants of integration. Furthermore, we obtain
\begin{equation}\label{s1_22}
\epsilon_2=\frac{\dot{\epsilon}_1}{H\epsilon_1}=\frac{1}{\epsilon_1}\frac{d\epsilon_1}{d\cal{N}}=\frac{-6}{1+\frac{C_2}{C_1}e^{6\cal{N}}}\quad\text{(close to $(0,0)$)}.
\end{equation}
Since ${\cal{N}}=\int H dt$ is the number of $e$-folds of inflationary expansion, one can choose $C_1$ and $C_2$ in such a way that $\epsilon_1\ll1$ and $\epsilon_2\ll1$ during inflation. For this goal, it is sufficient to take
\begin{equation}\label{sb1}
C_1,C_2\ll1,\quad\frac{C_2}{C_1}\gg1.
\end{equation}

Just to complete our knowledge about Eq. (\ref{s1_17}), let us consider the other fixed point in Eq. (\ref{s1_18}), i.e. $(2,-2)$.  The linearized system around $(2,-2)$ is
\begin{equation}
\begin{pmatrix}
\frac{d\epsilon_1}{d\cal{N}}&   \\
\frac{dy}{d\cal{N}} &
\end{pmatrix}=A_{**}\begin{pmatrix}
\epsilon_1-1 &   \\
y-1 &
\end{pmatrix},
\end{equation}
where
\begin{equation}
A_{**}=\begin{pmatrix}
8 & 1  \\
-4(3+g) & 0
\end{pmatrix}.
\end{equation}
$A_{**}$ has two positive eigenvalues as $\lambda_1=4+2\sqrt{1-g}$  and $\lambda_2=4-2\sqrt{1-g}$. Thus, for any value for $g$, one of the eigenvalues of $A_{**}$ is positive. Therefore $(2,-2)$ is not a stable fixed point.

\begin{figure}[ht]
 \includegraphics[width=9cm]{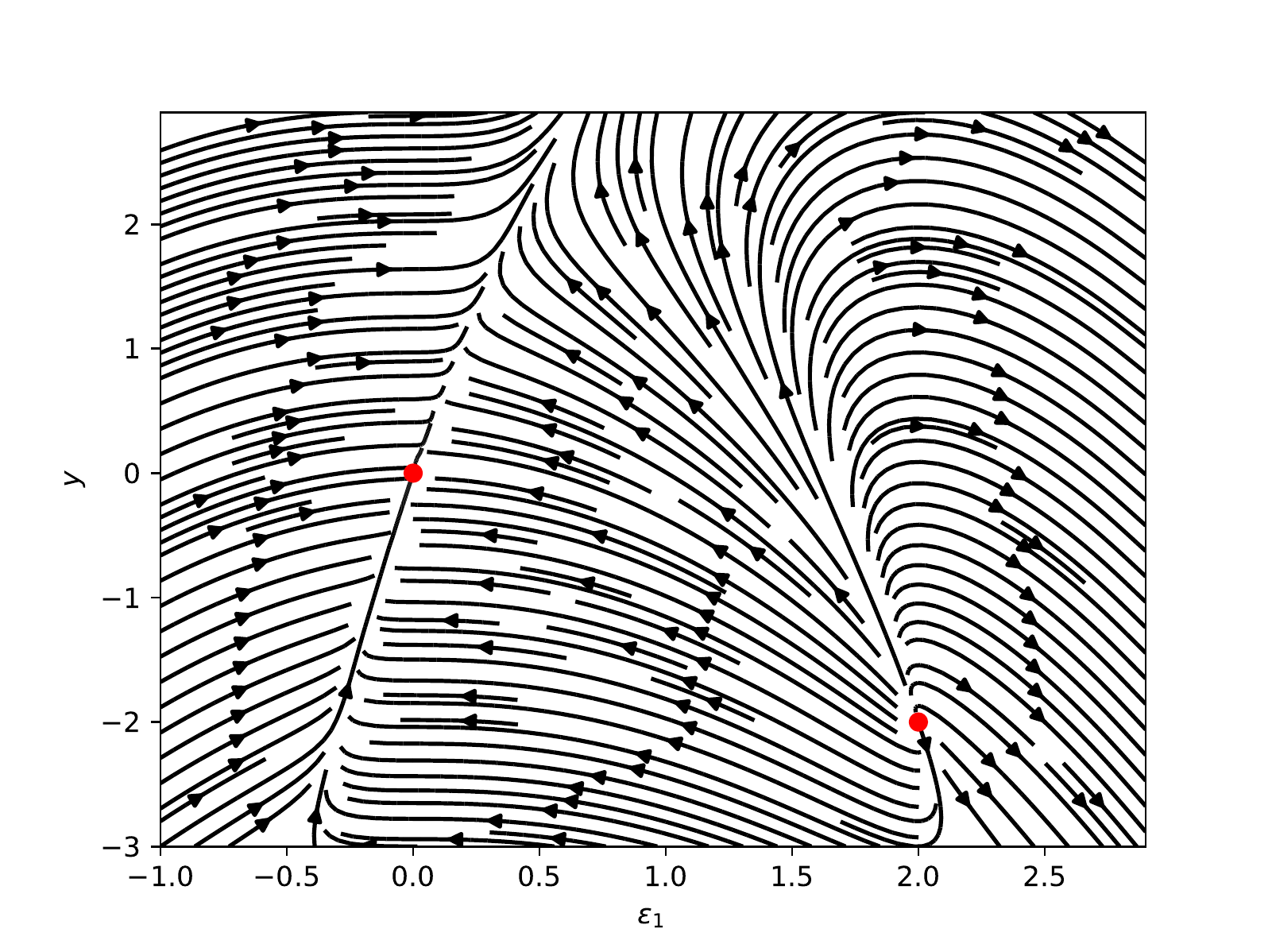}
 \caption{The phase space portrait of Eq. (\ref{s1_17}) for $g=0$. $(0,0)$ is the stable fixed point and $(2,-2)$ is the unstable fixed point.}
 \label{fig1}
\end{figure}

\begin{figure}[ht]
\includegraphics[width=9cm]{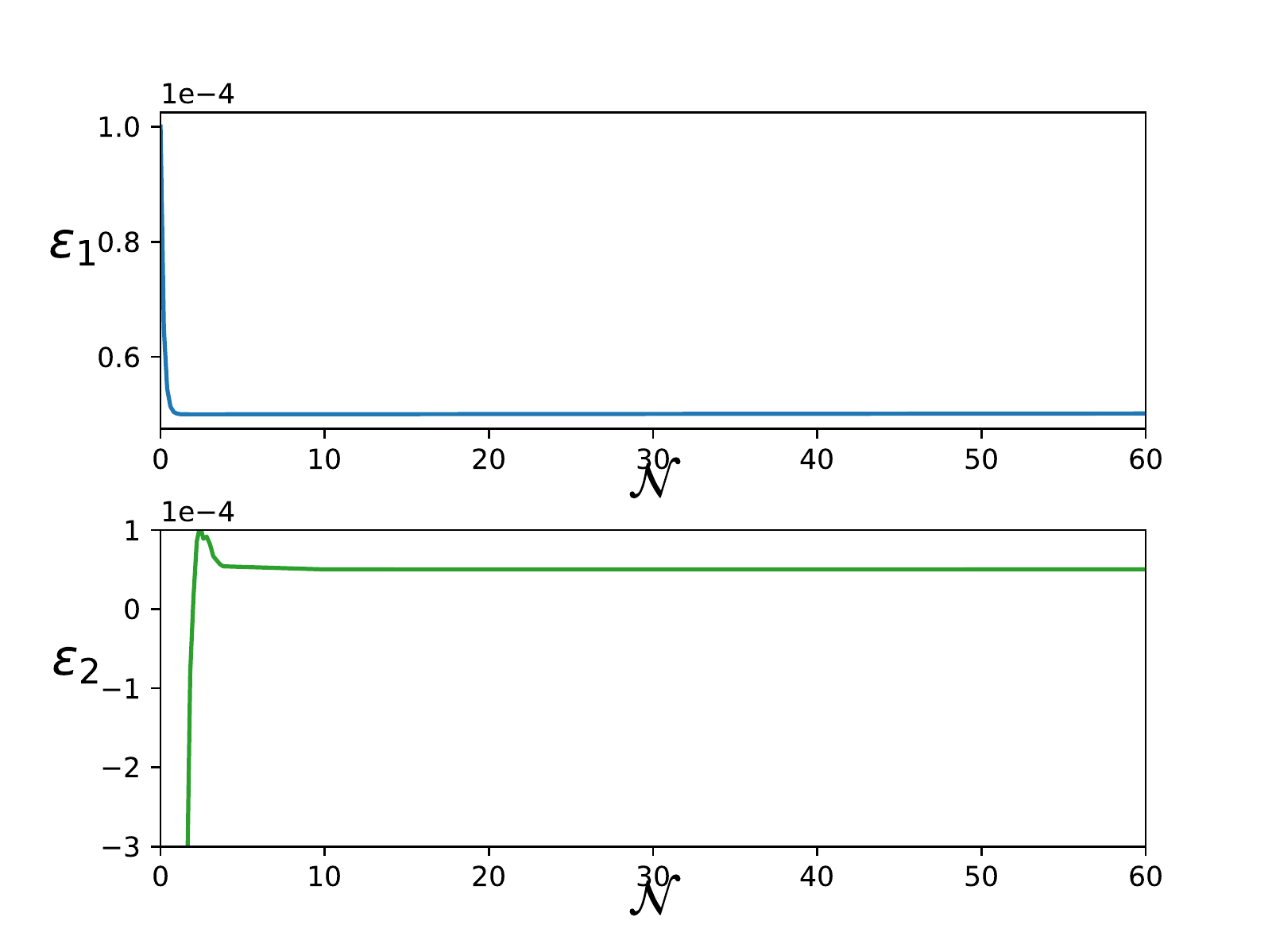}
\caption{$\epsilon_1$ and $\epsilon_2$ as functions of the number of $e-$foldings of the expansion during inflation $\mathcal{N}$. For these numerical solutions of Eq. \eqref{s1_16} we take $g=0$, and the initial values are $\epsilon_1=10^{-4}$ and $\epsilon_2=-3\times10^{-4}$. As we discuss in the main text, $\epsilon_1$ and $\epsilon_2$ remain small during inflation.}
\label{fig2}
\end{figure}

Fig.~\ref{fig1}, shows the phase space portrait of Eq. (\ref{s1_17}) for $g=0$, which is in agreement with our discussions. Numerical solutions confirm our results as is shown in Fig.~\ref{fig2}.
Here, the important result is that if we take initial conditions in such a way that $\epsilon_i\ll1$, the dynamics of the system do not refute such  assumptions.

%66666666666666666666666666666666

Recall that we have to consider Eq. (\ref{s1_af}) as the essential requirement that leads us to the stated results.
Eventually, at some time $t=t_f$, the Hubble parameter has decreased so much that the assumption made in Eq. (\ref{s1_af}) is longer valid. We want to determine the maximum number of \textit{e}-folds ${{\cal{N}}_{max}}$ for which Eq. (\ref{s1_af}) is valid.
Consider $t_i$ as the time in the beginning of inflation
and $t_f$ as the last time that Eq. (\ref{s1_af}) is valid.
Then Eq. (\ref{s1_22}) gives
\begin{equation}\label{sb2}
\epsilon_1(t_i)\approx C_1+C_2,\quad\epsilon_1(t_f)\approx C_2.
\end{equation}
On the other hand, since $\epsilon_1=-\frac{\dot{H}}{H^2}=-\frac{1}{H}\frac{dH}{d\cal{N}}$, Eq. (\ref{s1_22}) can be solved in terms of $N$ as
\begin{align}\label{Htf}
H(t_f)&=H(t_i)\l(e^{-\frac{C_1}{6}}e^{-C_2{\cal{N}}}\exp[C_2e^{-6\cal{N}}]\r)\\\nn
&\approx H(t_i)e^{-C_2{\cal{N}}}.
\end{align}
Since Eq.~\eqref{s1_af} is valid as far as $H(t_f)\approx H(t_i)$, according to Eq.~\eqref{Htf} we have to impose the following relation
\begin{equation}\label{sb3}
C_2 {{\cal{N}}_{max}} < \frac{1}{2}.
\end{equation}
Then, from Eqs. (\ref{sb1}), (\ref{sb2}) and (\ref{sb3}) it follows that
\begin{equation}\label{sb4}
{{\cal{N}}_{max}}< \frac{1}{2\epsilon_1(t_f)}\approx\frac{1}{2\epsilon_1(t_i)}.
\end{equation}
Therefore, if we take $\epsilon_1\approx 4\times 10^{-3}$, we have ${{\cal{N}}_{max}}<125$, which is sufficient to solve the horizon problem \cite{weinberg}. Recall that for the Starobinsky model the \textit{Planck} results \cite{Planck3} give the number of \textit{e}-folds to the end of inflation as $49<\mathcal{N}_{*}<59$ (95\% CL).

Returning to the dynamics of $X$, Eq. (\ref{s1_12}) can be solved as
\begin{align}\label{s1_23}
X-1=X_{*}\exp{\l(-3\int K(\epsilon_1,\epsilon_2){d\cal{N}}\r)},
\end{align}
where $X_{*}$  is a constant of integration. Eqs. (\ref{s1_13}), (\ref{s1_21}), (\ref{s1_22}) and (\ref{s1_23}) imply that as time passes $X$ evolves toward one, $X\rightarrow 1$. Thus, if choose an initial condition such that $X$ is close to one, it remains close to $X=1$ during inflation. Therefore, the scalar field is condensed during inflation. In the next section we will show that this fact leads to a nontrivial speed of sound for the scalar perturbations.

\section{Linear perturbations}\label{sec3}
In this section, we study the linear perturbations of the model. We have two objectives for this section. Firstly, we want to obtain observable quantities for the model, and secondly, we want to study potential ghost modes and tachyonic instability in the model. Table~\ref{table1} and Fig.~\ref{fig3} show the main results for the inflationary parameters. We will show that the lower bound on $f_s$ results to well-defined dynamics for the model. For this goal a suitable gauge is used. As discussed in Ref. \cite{{re},{rador}}, this is a nontrivial task for modified theories of gravity.
Then, we expand the action (\ref{i-1}) to the second order in scalar and tensor perturbations and study their implications for the model.
\subsection{Gauge fixing for the scalar perturbations}
In this part, we write the perturbed equations as
\begin{equation}\label{s2_1}
\delta G_{\mu}^{\nu}=\Mp^{-2}\delta T_{\mu}^{\nu},
\end{equation}
where $G_{\mu\nu}=R_{\mu\nu}-\frac{1}{2}Rg_{\mu\nu}$ is the Einstein tensor, and $\delta T_{\mu\nu}$ is corresponding  perturbed quantities.\\
To choose the gauge, let us parametrize the perturbed metric as \cite{weinberg}
\begin{align}\label{s2_2}
ds^2=&-\l(1+2\Phi(\vx,t)\r)dt^2+2a\pa_iB(\vx,t)dtdx^i\\\nn
&+\l(1+2\zeta(\vx,t)\r)a^2\delta_{ij}dx^idx^j.
\end{align}
Also, we define the following variables
\begin{align}\label{s2_3}
& \tilde{\Phi}(\vx,t)=\Phi(\vx,t)+\frac{2\Mp^{-2}}{1+4\Mp^{-2}Rf}\delta(Rf),\nn\\\
& \tilde{\zeta}(\vx,t)=\zeta(\vx,t)+\frac{2\Mp^{-2}}{1+4\Mp^{-2}Rf}\delta(Rf).
\end{align}
Using Eq. (\ref{s2_2}), the $(0i)$-component of Eq. (\ref{s1_2}) can be obtained as (after dropping $\pa_i$ from both sides)
\begin{align}\label{s2_4}
\tilde{H}\tilde{\Phi}(\vx,t)-\dot{\tilde{\zeta}}=&\frac{2\Mp^{-2}}{1+4\Mp^{-2}Rf}\bigg[\frac{Xf_X R^2}{2\dot{\varphi}}\delta\varphi\\\nn
&+3(\tilde{H}-H)\delta(Rf)\bigg].
\end{align}
The $(ii)$-component of Eq. (\ref{s1_2}) is obtained using Eq. (\ref{s2_2}) as
\begin{align}\label{s2_5}
&(3\th^2+2\dot{\th})\tilde{\Phi}(\vx,t)+\th\dot{\tilde{\Phi}}(\vx,t)-\ddot{\tilde{\zeta}}(\vx,t)\\\nn
&-3\th\dot{\tilde{\zeta}}(\vx,t)=\bigg[\Mp^{2}(4H\th-3H^2-\th^2)\tilde{\Phi}(\vx,t)\\\nn
&+\big(8H\th+4\th^2-3\dot{H}-8H^2+8\dot{\th}\big)\delta(Rf)\\\nn
&-\frac{Rf\delta(R)}{2}+6(\th-H)\frac{d\delta(Rf)}{dt}\bigg]\frac{\Mp^{-2}}{1+4\Mp^{-2}Rf}\\\nn
&-(\th-H)\dot{\tilde{\zeta}}(\vx,t),
\end{align}
where Eq. (\ref{s1_6}) is used to obtain the above result.

As discussed in Ref. \cite{weinberg}, one can define the scalar component of the velocity perturbation, $\delta u$, as $\delta G^{0}_{i}=\Mp^{-2}(\bar{\rho}+\bar{p})\pa_i \delta u$. Then the comoving gauge is defined as the gauge in which $\delta u=0$. So, from Eq. (\ref{s2_4}), the comoving gauge can be imposed with the following condition
 \begin{equation}\label{s2_6}
\frac{Xf_X R^2}{2\dot{\varphi}}\delta\varphi+3(\tilde{H}-H)\delta(Rf)=0.
   \end{equation}
Therefore, in the comoving gauge, we have $\dot{\tilde{\zeta}}(\vx,t)=\th\tilde{\Phi}(\vx,t)$. Substituting this into the left-hand side (LHS) of Eq. (\ref{s2_5}) results in
   \begin{equation}\label{s2_7}
   \text{LHS of Eq. (\ref{s2_5})}= \dot{\tilde{H}}\tilde{\Phi}(\vx,t).
   \end{equation}
To obtain the right-hand side (RHS) of Eq. (\ref{s2_5}), note that
\begin{equation}\label{s2_8}
\delta(Rf)=Rf_X+f\delta^{1}R,\hspace{.1cm}\delta X=-2X\Phi(\vx,t)+2X\frac{\delta\dot{\varphi}}{\dot{\varphi}},
\end{equation}
where $\delta^1R$ is the linear part of the perturbed $R$.
So,
\begin{equation}\label{s2_9}
fR\delta^1R=R\delta(Rf)+2R^2Xf_X\l(\tilde{\Phi}(\vx,t)-\frac{\delta\dot{\varphi}}{\dot{\varphi}}\r).
\end{equation}
Then, by taking the time derivative of Eq. (\ref{s2_6}) and using Eq. (\ref{s1_9}), we have
\begin{align}\label{s2_10}
&-\frac{Xf_XR^2}{6}\frac{\delta\dot{\varphi}}{\dot{\varphi}}=(\dot{\th}-\dot{H})\delta(Rf)\\\nn
&+(\th-H)\frac{d}{dt}\delta(Rf)
+3H(\th-H)\delta(Rf).
\end{align}
By eliminating $\delta\dot{\varphi}$ from  Eq. (\ref{s2_9}) and Eq. (\ref{s2_10}), we have
\begin{align}\label{s2_11}
fR\delta^1R=&\big[R+36H(\th-H)+12(\dot{\th}-\dot{H})\big]\delta(Rf)\\\nn
&+2R^2Xf_X\Phi(\vx,t)+12(\th-H)\frac{d}{dt}\delta(Rf).
\end{align}
Now, using $\dot{\tilde{\zeta}}(\vx,t)=\th\tilde{\Phi}(\vx,t)$ and by inserting the above result into RHS of Eq. (\ref{s2_5}), we have
\begin{equation}\label{s2_12}
 \text{RHS of Eq. (\ref{s2_5})}=\dot{\th}\tilde{\Phi}(\vx,t)\\\nn
 +2\Mp^{-2}(H^2+\dot{H})\delta(Rf).
\end{equation}
Comparing the above result with Eqs. (\ref{s2_7}) and (\ref{s2_6}) gives the following results
\begin{equation}\label{s2_13}
\delta(Rf)=0,\quad \delta\varphi=0.
\end{equation}
So, one can use Eq. (\ref{s2_13}) as the comoving gauge in our model.

Regarding the definitions of $\tilde{\zeta}(\vx,t)$, $\tilde{\Phi}(\vx,t)$
and Eq. (\ref{s2_13}), we will use $\zeta(\vx,t)$ and $\Phi(\vx,t)$ in the following sections.
%%%%%%%%%%%%%%%%%%%%%%%%%%%%%%%%%%%%%%%%%%%%%%%%%%%%%%%%%%%%%%%%%%%%%%%
\subsection{Linear scalar perturbations from the second order action}
In the comoving gauge, $\zeta(\vx,t)$ is a gauge invariant quantity \cite{weinberg}. So it is convenient to obtain a differential equation in terms of $\zeta(\vx,t)$. Also, note that we have shown that it is possible to use a gauge in which $\delta\varphi=0$. So, at the background level, this property enables us to set $X_*=1$.

One way to obtain the equation for $\zeta(\vx,t)$ is to use the perturbed $(00)$-component of Eq. (\ref{s1_2}). Using Eqs. (\ref{s1_3}) (\ref{s1_6}) and (\ref{s2_12}) the perturbed $(00)$-component of Eq. (\ref{s1_2}) is
\begin{equation}\label{s2_14}
\tilde{H}a\frac{\pa^2B(\vx,t)}{a^2}+\frac{\pa^2\zeta(\vx,t)}{a^2}
=\frac{a^2(1+4\Mp^{-2}Rf)}{\th^2}\Sigma\dot{\zeta}(\vx,t),
\end{equation}
where
\begin{equation}\label{s2_15}
\Sigma\equiv 3(\th-H)^2+\frac{2X\Mp^{-2}R^2}{1+4\Mp^{-2}Rf}\left[\frac{f_X}{2}+Xf_{XX}
-\frac{2Xf_X^2}{f}\right].
\end{equation}
From Eq. (\ref{s2_14}) and $\dot{\zeta}(\vx,t)=\th\Phi(\vx,t)$, it is easy to obtain the equation for $\zeta(\vx,t)$, as is shown by Eq. (\ref{s2_27}).
However, we need the second order action in the next section. Thus, we shall obtain the differential equation for $\zeta(\vx,t)$ by using the second order action.

For this purpose, we use the
Arnowitt-Deser-Misner (ADM) formalism \cite{arnowitt-1960}. In the ADM formalism, the spacetime is foliated by spacelike hypersurfaces with a unit normal vector $n^\alpha$, and
the metric is parametrized as~\cite{{arnowitt-1960}}
\begin{equation}\label{s2_16}
\d s^2 = -N^2dt^2
+ h_{ij} \l(N^{i}dt
+ \d x^{i}\r) \l(N^{j}dt +\d x^{j}\r),
\end{equation}
where~$N$, $N^{i}$, $h_{ij}$ are the lapse, the shift vector and the spatial metric respectably. Also, in the ADM formalism we have the following relations~\cite{{arnowitt-1960}}
\begin{align}\label{s2_17}
\sqrt{-g}&=N\sqrt{h},\\\nn
R&=\sR+K^{ij}K_{ij}-K^2-2\big(n^{\alpha}_{;\beta}n^{\beta}-n^{\alpha}n^{\beta}_{;\beta}\big)_{;\alpha},
\end{align}
where $;$ denotes the covariant derivative, $\sR$
is the Ricci scalar constructed from $h_{ij}$ and the extrinsic curvature of the spatial hypersurface is defined as
\begin{equation} \label{s2_18}
K_{ij} =\frac{1}{2N}\l[\dot{h}_{ij}
- \l({}^{(3)}\nabla_i\, N_j + {}^{(3)}\nabla_j\, N_i\r)\r],
\end{equation}
where $K = h_{ij}K^{ij}$ and ${}^{(3)}\nabla$ denotes the covariant derivative with respect to $h_{ij}$.

Following \cite{mald}, we will use $ h_{ij}=a^2e^{2\zeta(\vx,t)}\delta_{ij}$ and expand the shift and the lapse as
\begin{align}\label{s2_19}
& N=1+\alpha_1+\alpha_2+\ldots,\quad N_i=\pa_i\psi +\beta_i,\\\nn
&\psi=\psi_1+\psi_2+\ldots,\quad \beta_i=\beta_i^{(1)}+\beta_i^{(2)}+\ldots,
\end{align}
where $\pa_i\beta_i=0$.
Also, we will use the following expansion for the Ricci scalar
\begin{equation}\label{s2_20}
R=\bar{R}+\delta R=\bar{R}+\delta^1R+\delta^2R+\ldots,
 \end{equation}
where $\bar{R}$ is the background value of the Ricci scalar and  $i$ in $\delta^iR$ denotes the order of perturbations. The explicit form of the above quantities are given by Eqs. (\ref{apbb1}) and (\ref{apbb2}) in Appendix~\ref{appendix2}.

It follows from Eqs. (\ref{s2_2}), (\ref{s2_16}) and Eq. (\ref{s2_19}) that
\begin{equation}\label{s2_21}
\Phi(\vx,t)=\alpha_1,\quad aB(\vx,t)=\psi_1.
\end{equation}
The action (\ref{i-1}) is written up to second order in the perturbations as
\begin{align}\label{s2_22}
S= \frac{\Mp^2}{2}\int &\Big[a^3e^{3\zeta(\vx,t))}N\Big]\Big[\sR+K^{ij}K_{ij}-K^2\\\nn
&+2\Mp^{-2}f(X)\l(\bar{R}+\delta^1R+\delta^2R\r)^2\Big].
\end{align}
Now, using the comoving gauge $\dot{\zeta}(\vx,t)=\th\Phi(\vx,t)$ and Eq. (\ref{s2_13}), and using the fact that
\begin{equation}\label{s2_23}
\frac{1}{N^2}=1-2\frac{\dot{\zeta}(\vx,t)}{\th}-2\alpha_2+3\frac{\dot{\zeta}^2(\vx,t)}{\th^2}+\ldots,
\end{equation}
we have
\begin{align}\label{s2_24}
f(X)&=f(\bar{X})+\bar{X}\left(-2\frac{\dot{\zeta}(\vx,t)}{\th}-2\alpha_2
+3\frac{\dot{\zeta}^2(\vx,t)}{\th^2}\right)\\\nn
&\quad+2\bar{X}^2\frac{\dot{\zeta}^2(\vx,t)}{\th^2}+\ldots,
\end{align}
where $\bar{X}=\frac{\dot{\varphi}^2}{2M^4}$.
Plugging the above result into the action (\ref{s2_22}) and using Eqs. (\ref{s2_13}), (\ref{apbb1}), (\ref{apbb2}) and $\dot{\zeta}(\vx,t)=\th\Phi(\vx,t)$, leads us to the second order action, $\delta^2S$, as
\begin{align}\label{s2_25}
\delta^2S=\int & dt d^3\vx a^3\Mp^2\bigg[\frac{a^2\Sigma}{\th^2}\dot{\zeta}^2(\vx,t)\\\nn
&+(1+4\Mp^{-2}Rf)\bigg(\frac{\dot{\th}}{\th^2}+\frac{H}{\th}-1\bigg)\frac{(\pa\zeta(\vx,t))^2}{a^2}\bigg],
\end{align}
where we have used integration by parts and dropped surface terms.
Note that one can use Eq. (\ref{s1_6}) to find that
\begin{equation}\label{s2_26}
 \th-H-\frac{\dot{\th}}{\th}=\frac{1}{\th}\l(3(\th-H)^2+\frac{\Mp^{-2} Xf_XR^2}{1+\Mp^{-2}Rf}\r).
\end{equation}

Variation of Eq. (\ref{s2_25}) with respect to $\zeta(\vx,t)$, and using Eq. (\ref{s2_26}), gives
\begin{equation}\label{s2_27}
\ddot{\zeta}_k+\frac{\dot{w}_s}{w_s}\dot{\zeta}_k+\frac{k^2}{a^2}c_s^2\zeta_k=0,
\end{equation}
where
\begin{equation}\label{s2_28}
\zeta_k=\int d^3\vx\zeta(\vx,t)e^{-i\vk\cdot\vx},\quad
w_s\equiv a^3\frac{1+4\Mp^{-2}Rf}{\th^2}\Sigma,
\end{equation}
and $c_s$ is the speed of sound, which has the following form
\begin{align}\label{s2_29}
c_s^2&\equiv\frac{1}{1+\Xi},\\\nn \Xi&\equiv\frac{2M_p^{-2}X^2R^2\l(f_{XX}-2f_X^2\r)}{3(1+4\Mp^{-2}Rf)(\th-H)^2+\Mp^{-2}Xf_XR^2}.
\end{align}
From Eq. (\ref{s2_29}), it is clear that for the Starobinsky model, we have $c_s=1$. Note that at the minimum, $f_{X}=0$, $f_{XX}>0$, the above equation yields $\Xi>0$ and $0<c_s^2\leq1$. Also, Eq. (\ref{s2_29}) shows that higher order terms in the expansion  around the minimum of $f(X)$, are suppressed not only by powers of $(X-1)$ but also by $\frac{H}{\Mp}$. Since in Eq. (\ref{s2_29}) everything is exact, this property shows that the model is well-defined around the minimum of $f(X)$.

Let us next focus on the inflationary era. Expanding  Eq. (\ref{s2_29}) around the minimum of $f(X)$, i.e. $X=1$, and keeping the terms proportional to $\epsilon_1$ and $\epsilon_2$ results in
\begin{equation}\label{s2_30}
\Xi=\frac{f_{XX}}{f\epsilon_1^2}\l(2+2\epsilon_1-\epsilon_2\r)+\ldots,
\end{equation}
where Eq. (\ref{s1_af}) is used.
For reasons that will soon become clear, we have to impose the following condition
\begin{equation}
\frac{f_{XX}}{f\epsilon_1^2}\ll1.
\end{equation}
Thus, $\Xi\ll1$. Therefore, during inflation, we have
\begin{equation}\label{s2_31}
c_s^2=1-\frac{f_{XX}}{f\epsilon_1^2}\l(2+2\epsilon_1-\epsilon_2\r).
\end{equation}
Furthermore, it is convenient to define the following variable
\begin{equation}
s=\frac{\dot{c}_s}{Hc_s}.
\end{equation}
It follows from Eq. (\ref{s2_31}) that
 \begin{equation}
s\approx2\epsilon_2\frac{f_{XX}}{f\epsilon_1^2}\ll1.
\end{equation}
For fluctuations outside the horizon, $\frac{k}{aH}\ll1$, Eq. (\ref{s2_27}) simplifies as
\begin{equation}\label{s2_32}
\ddot{\zeta}_k+\frac{\dot{w}_s}{w_s}\dot{\zeta}_k\approx0,
\end{equation}
which has two solutions as
\begin{equation}\label{s_33}
\zeta_k=C_3,\quad\zeta_k=C_4\int\frac{dt}{w_s},
\end{equation}
where $C_3$ and $C_4$ are constants of integration.
From Eq. (\ref{s2_15}) it is clear that around the minimum of $f(X)$ we have $\Sigma>0$. If we impose the following condition
\begin{equation}
 1+4\Mp^{-2}Rf>0,
 \end{equation}
then it follows from Eq. (\ref{s2_28}) that $w_s>0$. Note that the above condition is consistent with Eq. (\ref{s1_af}).
Therefore, one of the solutions (\ref{s_33}) of Eq. (\ref{s2_32}) is a constant and the other solution decays.

In order to obtain the power spectrum of $\zeta_k$, it is convenient to represent Eq. (\ref{s2_25}) in the canonical form. For this purpose, we define the following variable
\begin{equation}\label{s2_34}
z^2\equiv\frac{2a^2\l(1+4\Mp^{-2}Rf\r)}{\th^2}\Sigma,
\end{equation}
and the conformal time as $d\tau=\frac{dt}{a}$. Then Eq. (\ref{s2_25}) takes the following form
\begin{align}\label{s2_35}
\delta^2S=\Mp^2 &\int d^3xd\tau\bigg[\frac{z^2}{2}(\zeta^\prime(\vx,\tau))^2\\\nn
&+a^2(1+4\Mp^{-2}Rf)\bigg(\frac{\dot{\th}}{\th^2}+\frac{H}{\th}-1\bigg)(\pa\zeta(\vx,\tau))^2\bigg],
\end{align}
where a prime $^\prime$  denotes a derivative with respect to $\tau$. Then variation of Eq. (\ref{s2_35}) with respect to $\zeta(\vx,\tau)$ and using the corresponding Fourier component $\zeta_k$, yields
\begin{equation}\label{s2_36}
\zeta^{\prime\prime}_k+2\frac{z^\prime}{z}\zeta^{\prime}_k+c_s^2k^2\zeta_k=0.
\end{equation}
Using Eqs. (\ref{s2_15}), (\ref{s2_26}), (\ref{s2_34}), and imposing Eq. (\ref{s1_af}), gives
\begin{equation}\label{s2_37}
\zeta^{\prime\prime}_k+2aH\l(1+\frac{\epsilon_2}{2}-s\r)
 \zeta^{\prime}_k+c_s^2k^2\zeta_k=0,
\end{equation}
where we keep the terms proportional to $\epsilon_i$.
Then we use the so-called slow-roll approximation which asserts that during inflation we have \cite{weinberg}
\begin{equation}\label{s2_38}
aH\approx\frac{-1}{(1-\epsilon_1)\tau}.
\end{equation}
Inserting Eq. (\ref{s2_38}) into Eq. (\ref{s2_37}), and keeping only terms of the first order in the perturbations, results in
\begin{equation}\label{s2_39}
\zeta^{\prime\prime}_k-\frac{2}{\tau}\l(1+\epsilon_1-s+\frac{\epsilon_2}{2}\r)
 \zeta^{\prime}_k+c_s^2k^2\zeta_k=0.
\end{equation}
In order to solve the above equation, consider a new variable $v_k\equiv \Mp z\zeta_k$, and then using Eq. (\ref{s2_34}) rewrite Eq.~\eqref{s2_39} as
\begin{equation}\label{s2_40}
v^{\prime\prime}_k+\l(c_s^2k^2-\frac{\nu_s^2-\frac{1}{4}}{\tau^2}\r)v_k=0,
\end{equation}
where
\begin{equation}\label{s2_41}
\nu_s\equiv\frac{3}{2}+\epsilon_1+\frac{\epsilon_2}{2}-s.
\end{equation}
Eq. (\ref{s2_40}) has two solutions as $\sqrt{-\tau}H_{\nu_s}^{(1)}(-c_sk\tau)$ and
$\sqrt{-\tau}H_{\nu_s}^{(2)}(-c_sk\tau)$, where $H_{\nu_s}^{(1)}$ and $H_{\nu_s}^{(2)}$ are the Hankel functions for which we have $H_{\nu_s}^{(1)}=H_{\nu_s}^{(2)*}$.
To find suitable combinations of the solutions, note that the asymptotic expansion of the Hankel function has the following form \cite{hankel}
\begin{equation}\label{s2_42}
\lim_{k\tau\rightarrow-\infty}H_{\nu_s}^{(1)}(-c_sk\tau)
=\sqrt{\frac{2}{\pi}}\frac{1}{\sqrt{-c_sk\tau}}e^{-ic_sk\tau}e^{-i\frac{\pi}{2}(\nu_s+\frac{1}{2})}.
\end{equation}
If we take the Bunch-Davies vacuum for the perturbations deep inside the horizon, i.e. ${k\tau\rightarrow-\infty}$, the suitable solution is \cite{vac}
\begin{equation}\label{s2_43}
v_k=\sqrt{\frac{\pi}{2}}e^{i\frac{\pi}{2}(\nu_s+\frac{1}{2})}\sqrt{-\tau}H_{\nu_s}^{(1)}(-c_sk\tau).
\end{equation}
For the limit outside the horizon, i.e. ${k\tau\rightarrow 0}$,
Eq. (\ref{s2_43}) gives that \cite{hankel}
\begin{equation}\label{s2_44}
v_k\approx\frac{i}{\pi}\Gamma(\nu_s)\sqrt{\frac{\pi}{2}}e^{i\frac{\pi}{2}(\nu_s+\frac{1}{2})}\sqrt{-\tau}\l(\frac{-c_sk\tau}{2}\r)^{-\nu_s}.
\end{equation}
The power spectrum of $v_k$ is defined as
\begin{equation}\label{s2_45}
\langle v_\vk v_\vk' \rangle={P_{v}}\delta(\vk+\vk').
\end{equation}
Also, from $v_k=\Mp z\zeta_k$, it follows that the power spectrum of curvature perturbations, $P_{\zeta}$, can be obtained by  $P_{\zeta}=\Mp^{-2}z^{-2}P_{v}$. Also, the dimensionless scalar power spectrum, $\Delta^2_s$, is defined as
\begin{equation}
\Delta^2_s=\frac{k^3 P_{\zeta}}{2\pi^2}.
\end{equation}
So, from Eqs. (\ref{s2_34}), (\ref{s2_44}) and (\ref{s2_45}), we have
\begin{equation}\label{s2_46}
\Delta^2_s=\frac{\Mp^{-2}}{4\pi^2}k^3\Gamma^2(\nu_s)z^{-2}(-\tau)\l(\frac{-c_sk\tau}{2}\r)^{-2\nu_s}.
\end{equation}
Therefore, the scalar spectral index $n_s$ is obtained as
\begin{equation}\label{s2_47}
n_s-1=\frac{d\ln(\Delta^2_s)}{d\ln k}=3-2\nu_s,
\end{equation}
which gives
\begin{equation}\label{s2_48}
n_s-1=-2\epsilon_1-\epsilon_2+2s=-2\epsilon_1-\l(1-4\frac{f_{XX}}{f\epsilon_1^2}\r)\epsilon_2.
\end{equation}
If we take $f_{XX}=0$, the above result reduces to the corresponding result for the Starobinsky model.
Note that for the Starobinsky model in the Einstein frame, we have $\epsilon_2=2\epsilon_1$ \cite{re}.

The \textit{Planck} results are in good agreement with the Starobinsky model. So, we have to impose $4\frac{f_{XX}}{f\epsilon_1^2}<10^{-2}$. Using $\epsilon_1\approx 4\times10^{-3}$, it turns out that
\begin{equation}
\frac{f_{XX}}{f}<4\times10^{-8}.
\end{equation}
This condition is satisfied if we take $f(X=1)\geq 3\times10^{9}$, and $f_{XX}\approx {{\cal{O}}(1)}$.

\subsection{Tensor perturbations from the second order action}
For the tensor perturbation, we consider the following metric \cite{weinberg}
\begin{equation}\label{s2_49}
ds^2=-dt^2+a^2(\delta_{ij}+h_{ij}(\vx,t))dx^idx^j,
\end{equation}
where $h_i^i=\pa_ih_{ij}=0$.
The tensor perturbations have two polarization modes $(+,\times)$ and the Fourier representation is written as \cite{weinberg}
\begin{equation}\label{s2_50}
 h_{ij}(\vx,t)=\int \frac{d^3\vk}{(2\pi)^{\frac{3}{2}}}\sum_{\alpha=+,\times}\eta_{ij}^{\alpha}h_{\vk,\alpha}(t)e^{i\vk\cdot\vx},
\end{equation}
where $\eta^{\alpha}_{ij}\eta^{\alpha'}_{ij}=2\delta_{\alpha\alpha'}$ and $\eta^{\alpha}_{ii}=k^{i}\eta^{\alpha}_{ij}=0$.

We insert Eq. (\ref{s2_49}) into the action (\ref{i-1}), and use Eqs. (\ref{s1_3}) and (\ref{s1_6}), which gives the second order action for tensor perturbations as
\begin{equation}\label{s2_51}
\delta^2S_T=\frac{\Mp^2}{8}\int dtd^3\vx a^3(1+4\Mp^{-2}Rf)
\left[ \dot{h}^2_{ij}(\vx,t)
-\frac{(\pa h_{ij}(\vx,t))^2}{a^2}\right],
\end{equation}
where we have used integration by parts, dropped surface terms and the subscript $T$ stands for ``tensor''.
Variation of Eq. (\ref{s2_51}) with respect to $h_{ij}(\vx,t)$, and then using Eq. (\ref{s2_50}), yields
\begin{equation}\label{s2_52}
\ddot{h}_{\vk,\alpha}+\frac{\dot{w_T}}{w_T}\dot{h}_{\vk,\alpha}+k^2h_{\vk,\alpha}=0,
\end{equation}
where
\begin{equation}
w_T\equiv 1+4\Mp^{-2}Rf.
\end{equation}
For fluctuations outside the horizon, $\frac{k}{aH}\ll1$, Eq. (\ref{s2_52}) simplifies as
\begin{equation}\label{s2_53}
\ddot{h}_{\vk,\alpha}+\frac{\dot{w_T}}{w_T}\dot{h}_{\vk,\alpha}\approx0,
\end{equation}
which has two solutions as
\begin{equation}\label{s2_54}
h_{\vk,\alpha}=C_5,\quad h_{\vk,\alpha}=C_6\int\frac{dt}{w_T},
\end{equation}
where $C_5$ and $C_6$ are constants of integration. Since $w_T>0$, the above result shows that one solution is a constant while the other decays.

In order to obtain the power spectrum, we define the following variable
\begin{equation}\label{s2_55}
z_T^2\equiv\frac{a^2}{4}(1+4\Mp^{-2}Rf).
\end{equation}
Then the action (\ref{s2_51}) takes the following form,
\begin{equation}\label{s2_56}
\delta^2S_T=\frac{\Mp^2}{2}\int d\tau d^3x z_T^2(\l[(h'_{ij}(\vx,\tau))^2-(\pa h_{ij}(\vx,\tau))^2\r].
\end{equation}
Variation of Eq. (\ref{s2_56}) with respect to $h_{ij}(\vx,\tau)$
and using the corresponding Fourier component, $h_{\vk,\alpha}$, results in
\begin{equation}\label{s2_57}
 h_{\vk,\alpha}^{\prime\prime}+2\frac{z^\prime_T}{z_T}{h}_{\vk,\alpha}^{\prime}+k^2{h}_{\vk,\alpha}=0.
\end{equation}
Using  Eq. (\ref{s2_55}), and $v_{\vk,\alpha}\equiv\Mp z_Th_{\vk,\alpha}$, yields
\begin{equation}\label{s2_58}
v^{\prime\prime}_{\vk,\alpha}+\l[k^2-2a^2H^2(1-2\epsilon_1)\r]v_{\vk,\alpha}=0,
\end{equation}
where Eq. (\ref{s1_af}) is used and only terms of first order in perturbations have been kept.\\
Using Eqs. (\ref{s2_38}) and (\ref{s2_58}) then give, to first order in $\epsilon_i$,
\begin{equation}\label{s2_59}
v^{\prime\prime}_{\vk,\alpha}+(k^2-\frac{2}{\tau^2})v_{\vk,\alpha}=0.
\end{equation}
The above equation has two solutions as $\sqrt{-\tau}H_{\frac{3}{2}}^{(1)}(-k\tau)$, and $\sqrt{-\tau}H_{\frac{3}{2}}^{(2)}(-k\tau)$.
Using Eq. (\ref{s2_42})  and the Bunch-Davies vacuum, the suitable solution is obtained as
\begin{equation}\label{s2_60}
v_{\vk,\alpha}=\sqrt{\frac{\pi}{2}}e^{-i\pi}\sqrt{-\tau}H_{\frac{3}{2}}^{(1)}(-k\tau).
\end{equation}
Thus, for the limit outside the horizon, i.e. $k\tau\rightarrow0$, the above result gives
\begin{equation}\label{s2_61}
v_{\vk,\alpha}\approx\frac{i}{\pi}\Gamma(\frac{3}{2})\sqrt{\frac{\pi}{2}}e^{-i\pi}\sqrt{-\tau}\l(\frac{-k\tau}{2}\r)^{-\frac{3}{2}}.
\end{equation}
The power spectrum of $v_{\vk,\alpha}$, $P_{v_{\vk,\alpha}}$, is defined by
\begin{equation}\label{s2_62}
\langle v_{\vk,\alpha}v_{\vk^{\prime},\alpha} \rangle={P_{v_{\vk,\alpha}}}\delta(\vk+\vk^{\prime}).
\end{equation}
Since we have defined, $v_{\vk,\alpha}=\Mp z_T h_{\vk,\alpha}$, it follows that the power spectrum of $h_{\vk,\alpha}$ can be obtained by $P_{h_{\vk,\alpha}}=\Mp^{-2}z_T^{-2}P_{v_{\vk,\alpha}}$.
The dimensionless tensor power spectrum $\Delta^2_T$ is defined as
$\Delta^2_T=2\frac{k^3 P_{h_{\vk,\alpha}}}{2\pi^2}$, and hence
\begin{equation}\label{s2_63}
\Delta^2_T=k^3\frac{\Mp^{-2}}{\pi^2}z_T^{-2}P_{v_{\vk,\alpha}}.
\end{equation}
Thus, we obtain the dimensionless power spectrum of the tensor perturbations from Eqs. (\ref{s2_61}), (\ref{s2_62}) and (\ref{s2_63}) as
\begin{equation}\label{s2_64}
 \Delta^2_T=\frac{\Mp^{-2}}{2\pi^2}\Gamma^2({\frac{3}{2}})k^3{z_T^{-2}}(-\tau)\left(\frac{-k\tau}{2}\right)^{-3}.
\end{equation}
The spectral index for the tensor perturbations, $n_T$, is given by
\begin{equation}\label{s2_65}
n_T=\frac{d\ln (\Delta^2_T)}{d\ln k},
\end{equation}
which gives
\begin{equation}\label{report_te}
n_T=0  \hspace{0.3cm} (\text{in the leading order for $\Delta^2_T$)}.
\end{equation}
Note that, we use the Jordan frame. As is shown in  Ref. \cite{re}, the corresponding results for the Starobinsky model gives the same results in the Jordan frame.\footnote{For the Starobinsky model, it is easy to define the Einstein frame and then obtain $n_T$ in it. To see this point, in this footnote, consider $f=f_s$.
From Eqs. (\ref{s1_7}) and (\ref{s2_55}) it follows that $z_T=\frac{\tilde{a}}{2}$ and then Eq. (\ref{s2_57}) takes the following form
\begin{equation*}
 h_{\vk,\alpha}^{\prime\prime}+2\tilde{H}{h}_{\vk,\alpha}^{\prime}+k^2{h}_{\vk,\alpha}=0.
 \end{equation*}
This equation is the same as the equation for the tensor perturbations of a usual scalar filed in the Einstein frame.}
However, note that the last result is based on the fact that $\Delta^2_T$ does not depend on $k$ in the leading order.
As is shown in Ref. \cite{re21}, if one wants to check the consistency relation in a modified theory of gravity, sometimes it is necessary to  go beyond the leading order. We will see that in order to obtain the tensor-to-scalar ratio in our model (which is an important observable quantity), we just need to consider our results in the leading order.

\subsection{The tensor-to-scalar ratio}
The tensor-to-scalar ratio, $r$, is defined  by
\begin{equation}\label{s2_66}
r=\frac{\Delta^2_T}{\Delta^2_s}|_{k=Ha},
\end{equation}
where $k=Ha$ shows that we evaluate the tensor-to-scalar ratio
at the moment of horizon crossing.

Using Eqs. (\ref{s2_34}), (\ref{s2_38}) and (\ref{s2_46}), we have
\begin{equation}\label{s2_67}
 \Delta^2_s|_{k=Ha}=\frac{1}{12\pi^2(1+4\Mp^{-2}Rf)}\frac{c_s^{-1}}{\epsilon_1^2}\frac{H^2}{\Mp^2},
 \end{equation}
where we have set $\nu_s\approx \frac{3}{2}$ and used $\Gamma(\frac{3}{2})=\frac{\sqrt{\pi}}{2}$.
Using Eq. (\ref{s1_af}), the above result gives
\begin{equation}\label{s2_68}
 \Delta^2_s|_{k=Ha}=\frac{1}{288\pi^2f}\frac{c_s^{-1}}{\epsilon_1^2}.
 \end{equation}
Similarly, Eq. (\ref{s2_64}) gives
\begin{equation}\label{s2_69}
 \Delta^2_T|_{k=Ha}=\frac{4}{\pi^2(1+4\Mp^{-2}Rf)}\frac{H^2}{\Mp^2},
 \end{equation}
which leads us to the following result
\begin{equation}\label{s2_70}
 \Delta^2_T|_{k=Ha}=\frac{1}{6\pi^2f}.
\end{equation}
Finally, Eqs. (\ref{s2_66}), (\ref{s2_68}) and (\ref{s2_70}), result in
\begin{equation}\label{s2_71}
 r=48\epsilon_1^2c_s=48\epsilon_1^2\l(1-\frac{f_{XX}}{f\epsilon_1^2}\r),
 \end{equation}
where Eq. (\ref{s2_31}) is used.

\begin{figure}[ht]
 \includegraphics[width=10cm]{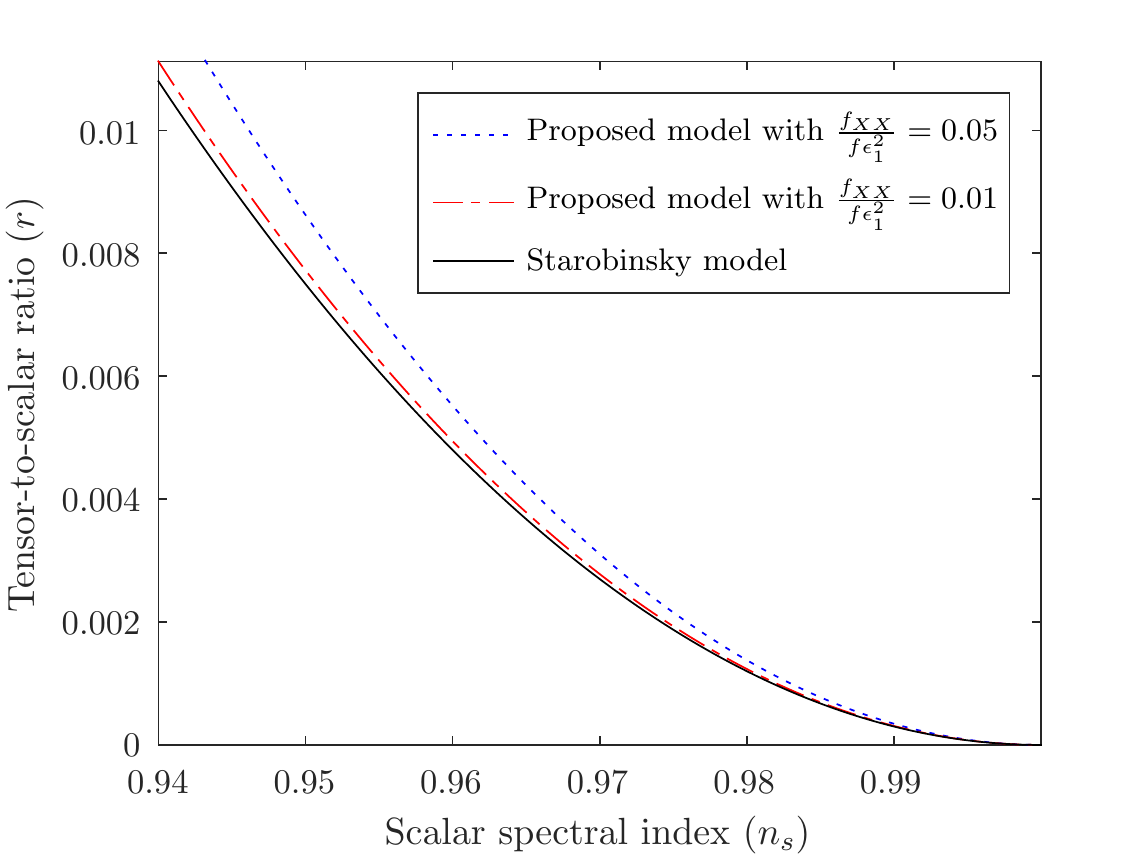}
 \caption{The predictions of the proposed model and the Starobinsky model in the $(n_s,r)$-plane. The Planck results \cite{Planck3} determine $n_s=0.9649\pm0.0042$ at $68\%$ CL and $r<0.056$ at $95\%$ CL. The predictions of the proposed model are very close to those of the Starobinsky model and in good agreement with the Planck results.}
 \label{fig3}
\end{figure}

The main results of this part are collected in Table~\ref{table1}, which are obtained from Eqs. \eqref{s2_48} and \eqref{s2_71} using the fact that
$\frac{f_{XX}}{f\epsilon_1^2}\ll1$. In Fig.~\ref{fig3}, we compare the proposed model with the Starobinsky model in the ($n_s,r)$-plane. As is shown in Fig.~\ref{fig3}, the predictions of the proposed model  are very close to the predictions of the Starobinsky model. Since observations give $\epsilon_1$ to be of the order of $10^{-3}$ (or slightly larger), and the two models have similar predictions if and only if $f\epsilon_1^2\geq 10^2$ (for a bound on $f_{XX}$ see below), having $f\geq 10^{9}$ ensures that the two models have similar predictions. Another approach to interpret the above results is that the Starobinsky model is just a special limit of the proposed model. From the theoretical side, the advantage of this approach is that it provides a reason for the value of $f_s$. From the observational side, it provides a model to compare with the Starobinsky model.
\begin{table}
\renewcommand{\arraystretch}{1.3}
    \caption{Comparing inflationary parameters.}
    \begin{tabular}{ | l | l | p{3.4cm}|}
    \hline
    Parameter & Starobinsky model & Proposed model  \\ \hline
    $n_s-1$ & $-4\epsilon_1$ & $-4\epsilon_1+\l(8\frac{f_{XX}}{f\epsilon_1^2}\r)\epsilon_1$  \\ \hline
    $n_T$ & 0 & 0  \\ \hline
    $r$ & $3(n_s-1)^2$ &$3(n_s-1)^2(1+3\frac{f_{XX}}{f\epsilon_1^2})$\\
    \hline
    \end{tabular}
    \label{table1}
\end{table}
%6666666666666666666666666666666666666666666666666666

\subsection{Absence of ghost  modes and tachyonic instability}
Since the model has extra degrees of freedom, in this part we will obtain conditions for which such extra degrees of freedom are not ghost and/or tachyons. We first focus on the scalar perturbations.
In order to avoid ghost modes in the model, the second order action in Eq. (\ref{s2_25}) gives the following conditions
\begin{equation}\label{report_1}
\Sigma>0,\quad-(1+4\Mp^{-2}Rf)\left(\frac{\dot{\th}}{\th^2}+\frac{H}{\th}-1\right)>0.
\end{equation}
These conditions ensure that the scalar perturbation is healthy in all stages of cosmological evolution.
Although one can use Eqs. (\ref{s2_15}), (\ref{s2_26}) and Eq. (\ref{report_1}) to impose conditions on the parameters of the model in any stage of cosmological evolution, we aim to impose the conditions generally on the function $f$. For this goal Eqs. (\ref{s2_15}), (\ref{s2_26})  and Eq. (\ref{report_1}) suggest the following conditions
\begin{equation}\label{report_2}
\frac{f_X}{2}+Xf_{XX}-\frac{2Xf_X^2}{f}\geq 0,\quad
1+4\Mp^{-2}Rf>0,\quad  f_X\geq0.
\end{equation}
These conditions must be satisfied during the whole cosmological evolution and regardless of the initial value of $X$ and the specific form of the function $f$. However, we do require that the initial value of $X$ is larger than the value of $X$ at the minimum of $f$, i.e., initially $X>1$, which ensures that $f_X\geq0$ can be satisfied, where the minimum of $f$ can be set to be at $X=1$ (by redefining the constant $M$).
It is interesting that in the vicinity of the minimum of $f$, where we can use $f=f_s+\frac{{\cal{F}}}{2}(X-1)^2+\ldots$, the conditions \eqref{report_2} are satisfied if $f_s\gg1$. Note that this condition is consistent with our motivations and goals in this paper. These conditions are also consistent with the absence of ghosts in the Starobinsky model.

In order to avoid tachyonic instability of the scalar perturbation, Eq. (\ref{s2_29}) gives the following conditions for well-defined behavior
\begin{equation}
f_{XX}-2f_X^2\geq 0,\quad f_X\geq 0.
\end{equation}
Thus, close to the minimum of $f$ it follows that $1-2{\cal{F}}(X-1)^2\geq0$, which shows that it is sufficient to require $f_{XX}={\cal{F}}\leq\frac{1}{2}$.

Regarding the tensor perturbations, we observe from Eq. (\ref{s2_56}) that well-defined behaviour is ensured when we impose $z_T^2\geq0$, which is satisfied by the conditions in Eq. (\ref{report_2}). Hence the healthiness of the tensor perturbations does not imply further conditions.
%55555555555555555555555555555555555555555555555555555

\section{The third-order action}\label{sec4}
The primordial non-Gaussianity is negligible for the Starobinsky model, which is in agreement with the observations \cite{{ng-re},{Planck9}}. So, it is necessary to study the action (\ref{i-1}) to obtain information about any additional sources for this sector.

The leading non-Gaussian signature arises from interactions in the third-order action. From the third-order action, one can obtain the three-point correlation function of the Fourier modes of the curvature perturbation which is related to the bi-spectrum, $B(k_1,k_2,k_3)$, as
\begin{align}
\langle\zeta_{\vk_{1}}\zeta_{\vk_{2}}\zeta_{\vk_3}\rangle=
(2\pi)^3\delta(\vk_1+\vk_2+\vk_3)B_\zeta(k_1,k_2,k_3).
\end{align}
Ref. \cite{Planck9} provides constraints on a dimensionless parameter, $f_{NL}$, which is defined as
\begin{equation}
f_{NL}(k_1,k_2,k_3)\equiv\frac{5}{6}\frac{B_\zeta(k_1,k_2,k_3)}{P_\zeta(k_1)P_\zeta(k_2)+2\text{ perms.}},
\end{equation}
where ``perms.'' stands for permutations.

In order to find the third-order action, we need $\delta^3R$ in Eq. (\ref{s2_20}), which is given by Eq. (\ref{apbb3}) in Appendix~\ref{appendix2}.
Using the comoving gauge, integrating by parts and the background equations of motion, the third-order action can be obtained as
\begin{align}
\delta^3S=&\int dt d^3xa^3\Bigg(\frac{\Mp^2}{2}[1+4\Mp^{-2}Rf]\\\nn
 &\times\Bigg[
\frac{2}{a^2}\zeta(\vx,t)(\pa\zeta(\vx,t))^2(\frac{\dot{\th}}{\th^2}
+\frac{H}{\th}-1)\\\nn
&+\frac{\dot{\zeta}(\vx,t)}{\th a^4}(\pa^2\psi_1)^2
-\frac{3}{a^4}\zeta(\vx,t)(\pa^2\psi_1)^2\\\nn
&-\frac{\dot{\zeta}(\vx,t)}{\th a^4}(\pa_i\pa_j\psi_1)(\pa_i\pa_j\psi_1)
-\frac{4}{a^4}\pa^2\psi_1\pa_i\zeta(\vx,t)\pa_i\psi_1\\\nn
&+\frac{3}{a^4}\zeta(\vx,t)(\pa_i\pa_j\psi_1)(\pa_i\pa_j\psi_1)
-2\frac{\dot{\zeta}^3(\vx,t)}{\th^3}\Sigma\\\nn
&+6\frac{\zeta(\vx,t)\dot{\zeta}^2(\vx,t)}{\th^2}\Sigma
\Bigg]\\\nn
&-12X^2R^2\frac{f_X^2}{f}\frac{\zeta(\vx,t)\dot{\zeta}^2(\vx,t)}{\th^2}\\\nn
&+4X^3R^2\Big(2\frac{f_Xf_{XX}}{f}-2\frac{f_X^3}{f^2}-\frac{f_{XXX}}{3}-\frac{f_{XX}}{2X}\Big)\frac{\dot{\zeta}^3(\vx,t)}{\th^3}\Bigg).
\end{align}
At $X=1$, the above action takes the following form
\begin{equation}\label{s5_4}
\delta^3S=\delta^3S|_{R^2}+f{\cal{S}}_f,
\end{equation}
where $\delta^3S|_{R^2}$ is the corresponding action for the Starobinsky model and
\begin{equation}
{\cal{S}}_f\equiv \frac{f_{XX}}{f}\int dt d^3x 2a^3R^2\l[3\frac{\dot{\zeta}^2(\vx,t){\zeta}(\vx,t)}{\th^2}-2\frac{\dot{\zeta}^3(\vx,t)}{\th^3}\r].
\end{equation}
In Eq. (\ref{s5_4}), $f$ is intentionally considered as the coefficient of ${\cal{S}}_f$.
Therefore, in the extended model we have two additional sources for the primordial non-Gaussianity at the minimum.
However, during inflation, $\delta^3S|_{R^2}$ is proportional to $f$, while ${\cal{S}}_f$ is suppressed by $\frac{f_{XX}}{f}$.

Using ${\cal{S}}_f$, it is straightforward to calculate the bi-spectrum as in Refs. \cite{{ng-re},{see}}. However, even for the Starobinsky model the primordial non-Gaussianity is small enough that we neglect it. Thus, regarding ${\cal{S}}_f$, which is further suppressed by $\frac{f_{XX}}{f}$ in comparison to $\delta^3S|_{R^2}$, we think that such calculations are not necessary for the proposed extension of the model.

\section{Conclusions}\label{sec5}
We have considered a model for the early Universe, which is motivated by the existence of a dimensionless constant in the Starobinsky model. Usually, we would prefer to have a mechanism to justify the existence of a very small or a very large value for a dimensionless constant in a model. In the cosmological context, we have shown that the proposed model has a mechanism from which the Starobinsky model of inflation emerges. Furthermore, the lower bound on the value of the constant parameter in the Starobinsky model can be justified by the dynamics of the proposed model. We have also obtained observable quantities of the proposed model by using cosmological perturbations.

The proposed model contains a scalar field, which is condensed during the inflation. We have shown that the absence of ghost modes leads us to a lower bound on the constant in the Starobinsky model. We have also studied the primordial non-Gaussianity in the model. Although there exist two additional sources of primordial non-Gaussianity in the proposed model, they are negligible if we impose now the obtained lower bound on the value of the constant in the Starobinsky model.

All our results lead us to conclude that the Starobinsky model emerges from the proposed model.
This conclusion is valuable if we compare it to other ways to generalize the Starobinsky model. i.e. by adding additional terms in the Jordan or Einstein frame.

Note that we have not claimed that the proposed model is a fundamental one. We just wish to introduce an alternative approach to interpret the Starobinsky model.

The implications of the proposed model for the reheating period and the study  of possible interactions of the scalar field with other particles are beyond the scope of this work and we intend to investigate them in future works.

% \section*{Acknowledgements}
\begin{acknowledgments}
We thank David Benisty, Stanley Deser, Dumitru Ghilencea, Tiberiu Harko, Sergei Ketov, Claus Montonen, Viatcheslav Mukhanov, Misao Sasaki and Ilya Shapiro for many valuable remarks and discussions.
\end{acknowledgments}

\appendix
\section{The model in the Einstein frame}\label{appendix1}
In this appendix, we show that the two actions for our model in \eqref{i-1} and \eqref{reei0}  are conformally equivalent.
The equivalent action in the Einstein frame is obtained with a standard procedure \cite{re}. We introduce two auxiliary fields $\Lambda$, $\Delta$, and then rewrite the action \eqref{i-1} as
\begin{align}\label{reap1}
S=\int d^4x\frac{\Mp^2}{2}\sqrt{-g}\left(\Lambda+\frac{2f(X)}{\Mp^2}\Lambda^2+\Delta(R-\Lambda)\right).
\end{align}
From variation of the above action with respect to $\Delta$, it is clear that \eqref{reap1} is equivalent to \eqref{i-1}. Also, variation of the above action with respect to $\Lambda$ results in
\begin{align}\label{reap2}
\Lambda=\frac{\Mp^2}{4f(X)}(\Delta-1).
\end{align}
Inserting Eq. \eqref{reap2} into \eqref{reap1} gives the following action
\begin{align}\label{reap3}
S=\int d^4x\frac{\Mp^2}{2}\sqrt{-g}\left(R\Delta-\frac{\Mp^2}{8f(X)}(1-\Delta)^2\right).
\end{align}

On the other hand, under the conformal transformation as $\tilde{g}_{\mu\nu}=\Omega^2 g_{\mu\nu}$ we have $\sqrt{-g}=\Omega^{-4}\sqrt{-\tilde{g}}$ and
\begin{align}\label{reap4}
R=\Omega^2\left(\tilde{R}+6\tilde{\nabla}_\mu(\tilde{g}^{\mu\nu}\frac{\partial_\nu\Omega}{\Omega})-6\tilde{g}^{\mu\nu}\frac{\partial_\mu\Omega\partial_\nu\Omega}{\Omega^2}\right),
\end{align}
where $\tilde{R}$ is the Ricci scalar constructed from $\tilde{g}_{\mu\nu}$ and the covariant derivative $\tilde{\nabla}$ is taken with respect to the metric $\tilde{g}_{\mu\nu}$.
Inserting Eq. \eqref{reap4} into \eqref{reap3} gives
\begin{equation}\label{reap5}
\begin{split}
S=\int d^4x\frac{\Mp^2}{2}\sqrt{-\tilde{g}}\Big(&\Omega^{-2}\tilde{R}\Delta+6\Omega^{-2}\Delta\tilde{\nabla}_\mu(\tilde{g}^{\mu\nu}\frac{\partial_\nu\Omega}{\Omega})-6\Omega^{-2}\Delta\tilde{g}^{\mu\nu}\frac{\partial\mu\Omega\partial_\nu\Omega}{\Omega^2}\\
&-\frac{\Mp^2\Omega^{-4}}{8f(\Omega^2\tilde{X})}(1-\Delta)^2\Big).
\end{split}
\end{equation}
where
\begin{align*}
\tilde{X}=\frac{-\tilde{g}^{\mu\nu}\partial_\mu\varphi\partial_\nu\varphi}{2M^4}=\frac{-\Omega^{-2}g^{\mu\nu}\partial_\mu\varphi\partial_\nu\varphi}{2M^4}=\Omega^{-2}X.
\end{align*}
Therefore, by choosing $\Omega^2=\Delta$ in \eqref{reap5} and dropping the surface term, it follows that
\begin{align}\label{reap6}
S=\int d^4x\frac{\Mp^2}{2}\sqrt{-\tilde{g}}\Big(\tilde{R}-\frac{3}{2}\tilde{g}^{\mu\nu}\frac{\partial_\mu\Delta\partial_\nu\Delta}{\Delta^2}
-\frac{\Mp^2\Delta^{-2}}{8f(\tilde{X}\Delta)}(1-\Delta)^2\Big).
\end{align}
Now, in order to have a canonical action, we define a scalar field $\phi$ as
\begin{align*}
\Delta=\exp\left(\frac{2\phi}{\sqrt{6}Mp}\right).
\end{align*}
Thus, \eqref{reap6} takes the following form
\begin{equation}\label{apreei01}
S_{E}=\int d^4x\sqrt{-\tilde{g}}\left[\frac{\Mp^2}{2} \tilde{R}-\frac{1}{2}\tilde{g}^{\mu\nu}\partial_\mu\phi\partial_\nu\phi-V(\phi,\varphi)\right],
\end{equation}
where
\begin{equation}\label{aprevisedei1}
V(\phi,\varphi)=\frac{\Mp^4}{16f\left(-\frac{1}{2M^4}e^{(\frac{2\phi}{\sqrt{6}Mp})}\tilde{g}^{\mu\nu}\partial_{\mu}\varphi\partial_\nu\varphi\right)}\left[1-\exp\left(\frac{-2\phi}{\sqrt{6}\Mp}\right)\right]^2.
\end{equation}
% \begin{equation}\label{aprevisedei1}
% V(\phi,\varphi)=\frac{\Mp^4}{16f\left(e^{(\frac{2\phi}{\sqrt{6}Mp})} \tilde{X}\right)}\left[1-\exp\left(\frac{-2\phi}{\sqrt{6}\Mp}\right)\right]^2.
% \end{equation}
Therefore, \eqref{i-1} and \eqref{reei0} are conformally equivalent.

% \begin{equation}
% \nabla_\mu j^\mu,\quad
% j^\mu=\sqrt{-\tilde g}\frac{\Mp^4 e^{(\frac{2\phi}{\sqrt{6}Mp})} f'\left(e^{(\frac{2\phi}{\sqrt{6}Mp})} \tilde{X}\right)}{16f\left(e^{(\frac{2\phi}{\sqrt{6}Mp})} \tilde{X}\right)^2}\left[1-\exp\left(\frac{-2\phi}{\sqrt{6}\Mp}\right)\right]^2
% \tilde{g}^{\mu\nu}\partial_\nu\varphi.
% \end{equation}

Note that \eqref{reei0} depends on two fields and since there is no symmetry that relates $\phi$ to $\varphi$, the study of perturbed equations and choosing a suitable gauge for \eqref{reei0} is not an easy task.

It is worth to mention that for the background cosmology, for which the FRW metric is used, $\phi$ and $\varphi$ depend only on time. Then \eqref{aprevisedei1} is reduced to a simple function for the FRW metric.
In this case, the potential is written as
 \begin{equation}\label{aprevisedei2}
V(\phi,\varphi)=\frac{\Mp^4}{16f\left(\frac{1}{2M^4}e^{(\frac{2\phi(\tilde{t})}{\sqrt{6}Mp})}\left(\frac{d\varphi}{d\tilde{t}}\right)^2\right)}\left[1-\exp\left(\frac{-2\phi(\tilde{t})}{\sqrt{6}\Mp}\right)\right]^2
\quad (\text{for the FRW metric}),
\end{equation}
where $\tilde{t}$ denotes time in the Einstein frame.
% Since $V(\phi,\varphi)$ is the only part of the Lagrangian where the field $\varphi$ appears, we can redefine that scalar field as $\tilde\varphi$ which satisfies $\frac{d\tilde\varphi}{d\tilde{t}}=e^{(\frac{\phi(\tilde{t})}{\sqrt{6}Mp})}\frac{d\varphi}{d\tilde{t}}$, i.e.,
% $\tilde\varphi(\tilde{t})-\tilde\varphi(\tilde{t}_0)=\int_{\tilde{t}_0}^{\tilde{t}}d\tau\, e^{(\frac{\phi(\tau)}{\sqrt{6}Mp})}\frac{d\varphi(\tau)}{d\tau}$,
% without affecting the rest of the action \eqref{apreei01}.
% Then we have
% Then, it is easy to show that under redefinition of time as $d\tilde{t}\rightarrow e^{(\frac{\phi(t)}{\sqrt{6}Mp})}d\tilde{t}$ the FRW metric does not change and we have
% \begin{equation}\label{aprevisedei3}
% V(\phi,\tilde\varphi)=\frac{\Mp^4}{16f\left(\frac{1}{2M^4}\left(\frac{d\tilde\varphi}{d\tilde{t}}\right)^2\right)}\left[1-\exp\left(\frac{-2\phi(\tilde{t})}{\sqrt{6}\Mp}\right)\right]^2
% \quad (\text{for the FRW metric}),
% \end{equation}
On the other hand, in Sec~\ref{sec2}, we have shown that $\varphi$ could be condensed in such way that $f\rightarrow f_s$. So, the above result shows that for the FRW metric we can have $V(\phi,\varphi)\rightarrow V(\phi)$ in the condensed phase. Therefore, the Starobinsky model emerges and we will have an inflation phase. Note that for the perturbed metric, the fields depend on $(\vec{x},t)$ and it is not possible to use the above arguments for the perturbed metric.

\section{The perturbed Ricci scalar}\label{appendix2}
In the main part of this paper, we have used the following expansion for the Ricci scalar
\begin{equation}
R=\bar{R}+\delta^1 R+\delta^2R+\delta^3 R+\ldots.
\end{equation}
Note that we have written the Ricci scalar in terms of the ADM variables 	as \cite{wald}
\begin{align}\label{s42_17000}
R=\sR+K^{ij}K_{ij}-K^2-2\big(n^{\alpha}_{;\beta}n^{\beta}-n^{\alpha}n^{\beta}_{;\beta}\big)_{;\alpha}.
\end{align}
The last term in the above formula is the total derivative term. So, this term has no effect in the Einstein-Hilbert action. However, in our work this term is important. Therefore, compared with other references, reader finds additional terms in the following formulas.

The explicit form of the above quantities are
\begin{equation}
\bar{R}=12H^2+6\dot{H},
\end{equation}
\begin{align}\label{apbb1}
\delta^1 R=&-24\alpha_1H^2+24H\dot{\zeta}-8H\pa_iN^i-6H\dot{\alpha_1}
\\\nn
&+6\ddot{\zeta}-2\pa_t\pa_iN^i-12\dot{H}\alpha_1-4a^{-2}\pa^2\zeta
\\\nn
&-2a^{-2}\pa^2\alpha_1,
\end{align}
\begin{align}\label{apbb2}
\delta^2 R=&-24H^2\alpha_2-2a^{-2}\pa^2\alpha_2-6H\dot{\alpha}_2-12\dot{H}
\alpha_2\\\nn
&+\frac{1}{2}\pa_iN^j\pa_iN^j
+8a^{-2}\zeta\pa^2\zeta-2a^{-2}\pa_i\zeta\pa_i\zeta\\\nn
&+4 a^{-2}
\zeta\pa^2\alpha_1
-2a^{-2}\pa_i\zeta\pa_i\alpha_1
+2a^{-2}\alpha_1\pa^2\alpha_1\\\nn
&+36\alpha_1^2H^2-48H\alpha_1\dot{\zeta}
+12\dot{\zeta}^2-6\dot{\alpha_1
}\dot{\zeta}\\\nn
&+18H\alpha_1\dot{\alpha}_1
-12\alpha_1\ddot{\zeta
}+18\alpha_1^2\dot{H}\\\nn
&-12\pa_i\dot{\zeta}N^i+16H\alpha_1\pa_iN^i
-24HN^i\pa_i\zeta\\\nn
&-8\dot{\zeta}\pa_iN^i
+2\dot{\alpha}_1\pa_iN^{i}-6\pa_{i}\zeta\pa_tN^{i}\\\nn
&+4\alpha_1\pa_t\pa_iN^{i}+6HN^i\pa_i\alpha_1+\pa_iN^i\pa_jN^j\\\nn
&+2N^i\pa_i(\pa_jN^j)+\frac{1}{2}\pa_iN^j\pa_jN^i,
\end{align}
\begin{align}\label{apbb3}
\delta^3R=&-8a^{-2}\zeta^2\pa^2\zeta+4a^{-2}\zeta\pa_i\zeta\pa_i\zeta+
72H\alpha_1^2\dot{\zeta}\\\nn
&-24H\alpha_1^2\pa_iN^i-24\alpha_1\dot{\zeta}^2
-\alpha_1\pa_iN^j
\pa_iN^j\\\nn
&-\alpha_1\pa_iN^j\pa_jN^i+48\alpha_1H\pa_i\zeta N^i+16\alpha_1\dot{\zeta}\pa_iN^i\\\nn
&+8\pa_iN^iN^j\pa_j\zeta-24\dot{\zeta}N^i\pa_i\zeta
-2\alpha_1\pa_iN^i\pa_jN^j\\\nn
&-4a^{-2}\zeta^2\pa^2\alpha_1+4\zeta
a^{-2}\pa_i\zeta\pa_i\alpha_1-4\zeta a^{-2}\alpha_1\pa^2\alpha_1\\\nn
&+4a^{-2}\alpha_1\pa_i\alpha_1\pa_i\alpha_1+2a^{-2}\pa_i\zeta\alpha_1
\pa_i\alpha_1+6\dot{\alpha}_1\pa_i\zeta N^i\\\nn
&+18\dot{\zeta}\dot{\alpha}_1
\alpha_1-6\alpha_1\dot{\alpha}_1\pa_iN^i
-36H\dot{\alpha}_1\alpha_1^2\\\nn
&+24\alpha_1\pa_i\dot{\zeta}N^i
+12\alpha_1\pa_i\zeta\pa_tN^i+18\ddot{\zeta}\alpha_1^2\\\nn
&-6\alpha_1^2\pa_t\pa_iN^i-18H\alpha_1\pa_i\alpha_1N^i
+6\dot{\zeta}N^i\pa_i\alpha_1\\\nn
&-2N^j\pa_iN^i\pa_j\alpha_1
+6N^i\pa_i(\pa_j\zeta N^j)\\\nn
&-4\alpha_1N^i\pa_i(\pa_jN^j)
-36H^2\alpha_1^3-24\dot{H}\alpha_1^3\\\nn
&+\text{ terms contain $\alpha_2$ and $\alpha_3$ }\\\nn
&+\pa_i(\text{ third-order terms}),
\end{align}
where $\zeta=\zeta(\vx,t)$.
%%%%%%%%%%%%%%%%%%%%%%%%%%%%%%%%%%%%%%%%%%%%%%%%%%%%%%%%%%%%%%
\bibliographystyle{apsrev4-1}
%\bibliography{biblio}

\begin{thebibliography}{99}
\bibitem{jordan}
We consider the Lagrangian \eqref{i-0} and the action \eqref{i-1} as models which are written in the physical frame (Jordan frame) in the sense that the baryons and the cold dark matter are minimally coupled to the metric $g_{\mu\nu}$ and obey the standard continuity equation \cite{re}. Note that, alternatively, one can consider \eqref{ei0} just as a given model without any relation to the $R^2$ model and consider it as the physical frame. Thus, the choice of the physical frame is a crucial part of the theory and it depends on the chosen point of view. During inflation, the effects of baryons and cold dark matter can be neglected.
\end{thebibliography}
%%%%%%%%%%%%%%%%%%%%%%%%%%%%%%%%%%%%%%%%%%%%%%%%%%%%%%%%%%%%%%
%\section*{References}

\end{document}